\documentclass[12pt,preprint]{aastex}

\shorttitle{{\em Spitzer} Observations of Cool WDs}
\shortauthors{Kilic et al.}

\begin{document}

\title{Spitzer Observations of the Oldest White Dwarfs in the Solar Neighborhood}

\author{Mukremin Kilic\altaffilmark{1}, Piotr M. Kowalski\altaffilmark{2}, William T. Reach\altaffilmark{3}, and Ted von Hippel\altaffilmark{4}}

\altaffiltext{1}{Spitzer Fellow, Smithsonian Astrophysical Observatory, 60 Garden Street, Cambridge, MA 02138;\\ mkilic@cfa.harvard.edu}
\altaffiltext{2}{Lehrstuhl f\"ur Theoretische Chemie, Ruhr-Universit\"at Bochum, 44780 Bochum, Germany}
\altaffiltext{3}{Spitzer Science Center, California Institute of Technology, Pasadena, CA 91125}
\altaffiltext{4}{Physics Department, Siena College, 515 Loudon Road, Loudonville, New York 12211}

\begin{abstract}

We present {\em Spitzer} 5-15 $\mu$m spectroscopy of one cool white dwarf
and 3.6 -- 8 $\mu$m photometry of 51 cool white dwarfs
with $T_{\rm eff}<6000$ K.
The majority of our targets have accurate BVRIJHK photometry and trigonometric parallax measurements available, which enables us to 
perform a detailed model atmosphere analysis using their optical, near- and mid-infrared photometry with
state-of-the-art model atmospheres. We demonstrate that the optical and infrared spectral energy distributions of cool white dwarfs
are well reproduced by our grid of models. Our best fit models are consistent with the observations within 5\% in {\it all} filters
except the IRAC 8 $\mu$m band, which has the lowest signal-to-noise ratio photometry. Excluding the ultracool white dwarfs,
none of the stars in our sample show significant mid-infrared flux deficits or excesses. The non-detection of mid-infrared excess
flux around our $\approx$2-9 Gyr old targets constrain the fraction of cool white dwarfs with warm debris disks to $0.8^{+1.5}_{-0.8}$ \%.

\end{abstract}


\section{Introduction}

Devoid of any nuclear burning, white dwarfs simply cool with time. This well understood cooling process \citep{mestel52} enables
astronomers to use them as cosmic chronometers. The absence of white dwarfs fainter than $\log (L/L_\odot)=-4.5$ in the Solar
Neighborhood is used as evidence of the finite age of the Galactic disk \citep{winget87}. The best estimate for the age of the
Galactic disk based on the white dwarf luminosity function is $8 \pm 1.5$ Gyr \citep{leggett98,harris06}.
This method has now been applied to two halo globular clusters, M4 and NGC 6397 \citep{hansen02,hansen07}.

Prior to the {\em Spitzer Space Telescope}, the oldest white dwarfs in the Solar Neighborhood could not be observed in the mid-infrared due to
their faintness. Therefore, the luminosities of cool white dwarfs were estimated using optical and near-infrared photometry, and
bolometric corrections based on white dwarf model atmospheres. An observational check on these corrections is required to
confirm their reliability. Such studies by \citet{kilic06a,tremblay07} and \citet{kilic08a} showed that the white dwarf models
are able to explain the spectral energy distributions (SEDs) of almost all cool white dwarfs with $T_{\rm eff}\geq 6000$ K. However,
significant mid-infrared flux deficits compared to model predictions were discovered for one star, LHS 1126.

The main atmospheric constituents in cool white dwarfs are hydrogen and helium, though trace amounts of heavier elements
are also observed in some white dwarfs. The atmospheric compositions of white dwarfs hotter than 5000 K can generally be
constrained based on the presence of Balmer lines.
However, Balmer lines disappear below about 5000 K, and optical spectroscopy cannot be used
to identify hydrogen versus helium rich white dwarfs. The most critical opacity sources in cool hydrogen-rich white dwarf
atmospheres are believed to be collision induced absorption (CIA) in the infrared
\citep{hansen98,saumon99} and Ly $\alpha$ in the ultraviolet \citep{kowalski06b, koester00}.
The primary opacity source in helium-rich white dwarfs
is He$^-$ free-free absorption \citep{kowalski07}.
Helium atmospheres are much denser than hydrogen atmospheres at the same temperature. Therefore if hydrogen is present,
collisions between H$_2$ molecules and neutral helium can induce CIA. This opacity becomes significant at higher temperatures
compared to pure hydrogen atmospheres.

CIA opacity is expected to produce molecular absorption features in the near-infrared, but
current CIA calculations do not predict absorption bands in the mid-infrared. Therefore, the observed mid-infrared flux deficits
for LHS 1126 imply that either the CIA calculations in white dwarf atmospheres are incomplete, or these flux deficits
are caused by some other mechanism \footnote{LHS 1126 shows blue shifted carbon bands in the optical \citep[see][]{hall08} and we cannot rule out
peculiarities in its infrared SED due to the presence of carbon or other elements.}.
An important question for white dwarf cosmochronology is whether LHS 1126 is unique
or not. The previous Spitzer observations of cool white dwarfs included only a few stars cooler than 6000 K, and
the mid-infrared flux distributions of the oldest white dwarfs in the Solar Neighborhood have not been studied yet.

In order to understand the mid-infrared SEDs of cool white dwarfs and the CIA opacity,
we used the Infrared Array Camera \citep[IRAC,][]{fazio04} on the {\em Spitzer Space Telescope} to observe 53 nearby cool white dwarfs.
We also used the Infrared Spectrograph (IRS) to observe one of our targets.
In this paper, we present IRAC photometry and IRS spectroscopy of these stars and perform detailed model atmosphere analysis
using both optical and
infrared data. Our observations and model fits are discussed in \S 2 and \S 3, respectively, while implications of these
data are discussed in \S 4 and \S 5.

\section{Sample Selection, Observations, and Reductions}

We selected our targets from the samples of \citet{bergeron97,bergeron01, bergeron05}. The beauty
of this sample selection is that Bergeron et al. provided accurate BVRIJHK photometry and trigonometric parallax measurements
for the majority of our targets. Our sample also includes three cool white dwarfs found in
the Sloan Digital Sky Survey (SDSS) by \citet{kilic06b}. These three stars have near-infrared photometry from the Two Micron
All Sky Survey \citep[2MASS,][]{cutri03}. We also added three ultracool white dwarfs, LHS 3250
\citep{harris99}, CE 51 \citep{ruiz01}, and WD 0346+246 \citep{hambly97}, to our sample to extend our study to $T_{\rm eff}<4000$ K.
The majority of our targets have accurate coordinates and proper motions measured by \citet{lepine05}, which is
very helpful in identifying these faint objects in crowded fields.

Observations reported here were obtained as part of our Cycle 3 GO-Program 30208 (PI: M. Kilic).
We obtained 3.6, 4.5, 5.8, and 8 $\mu$m images for 53 cool white dwarfs
including two common proper motion systems in 51 Astronomical Observation Requests (AORs).
Depending on the source brightnesses, integration times of 30 or 100 seconds per dither, with five or nine dithers for each target,
were used. We reduced the data using both IRAF and IDL routines. Our reduction procedures are similar to the
procedures employed by \citet{kilic08a} and \citet{mullally07}. Briefly, we use the IRAF PHOT and IDL astrolib packages to perform
aperture photometry on the individual BCD frames from the latest available IRAC pipeline reduction (S14.4.0, S15.3.0, S15.0.5,
or S16.1.0 for our targets). In order to maximize the signal-to-noise ratio, we used a 5 pixel aperture for bright, isolated
objects, and 2 or 3 pixel apertures for faint objects or objects in crowded fields.
Following the IRAC calibration procedure, corrections for the location of the source in the array
were taken into account before averaging the fluxes of each of the dithered frames at each wavelength.
Channel 1 (3.6$\mu$m) photometry was also corrected for the pixel-phase-dependence (see the IRAC Data Handbook).
The results from IRAF and IDL reductions were consistent within the errors. 
The photometric error bars were estimated from the observed scatter in the 5 (or 9) images (corresponding to the dither positions)
plus the 3\% absolute calibration error, added in quadrature. Finally, we divided the estimated fluxes by the color corrections for
a Rayleigh-Jeans spectrum \citep{reach05}. These corrections are 1.0111, 1.0121, 1.0155, and 1.0337 for the 3.6, 4.5, 5.8,
and 8 $\mu$m bands, respectively\footnote{These corrections were also used in \citet{kilic06a} and \citet{kilic08a}.}.

We present the average fluxes measured from the {\em Spitzer} images in Table 1. Two of the stars in our sample, WD 0346+246
and WD 1310$-$472 were blended with brighter stars, which prohibited us from performing photometry on these objects. These two
stars are not included in Table 1. Another common proper motion binary system, WD0727+482A and WD0727+482B, was unresolved
in IRAC images, and the photometry presented in Table 1 is the total flux from the system. We included this flux measurement
for completeness purposes, however we do not include these two stars in our analysis for the remainder of the paper.
Five other stars, WD 0222+648, WD 0551+468, WD 0851$-$246 (CE 51), WD 1247+550, and WD 2054$-$050, are near brighter stars in the IRAC
images, and the photometry is probably affected by these nearby sources. We present the photometry for these five stars and
include them in our analysis, but do not use them to draw any conclusions.

As part of our Cycle 3 program, we also used the Infrared Spectrograph \citep[IRS,][]{houck04}
Short-Low module in the first and second orders to obtain 5-15 $\mu$m spectroscopy
of the brightest cool white dwarf in our sample, 
Wolf 489 (WD 1334+039). These observations consisted of 44 cycles of 1 min exposures
and were performed on 2007 July 29. For each of the 44 basic calibrated spectral images
(BCDs) at each of the two nod positions and each of the two orders of the IRS,
a cube was compiled. Each cube was collapsed into a spectral image, using a robust average.
Outlier pixels with respect to a $5\times 5$ box median were replaced  
with the median, to account for rogue and dead pixels. The two nodded  
spectral images for each order were differenced. A significant spatial  
gradient is evident in the nod difference images, so we removed a row-by-row
(i.e. approximately constant wavelength) median spanning  
spectral order on the nod-difference image. The background-subtracted  
difference images were analyzed using SMART \citep{higdon04}, using  
manual extractions of tapered apertures and local background  
subtraction from the portion of the slit not covered by the source.  
The extracted spectra were merged and rebinned using a logarithmic bin  
spacing, yielding the final spectrum. 
 
\section{Model Atmospheres and the Fitting Procedures}

We use state of the art white dwarf model atmospheres to fit the BVRIJHK and {\em Spitzer} photometry of our targets.
Our fitting procedure is the same as the method employed by \citet{tremblay07}.
Briefly, the magnitudes are converted into monochromatic fluxes using the zero points derived from
the Vega (STIS) spectrum integrated over the passband for each filter. The resulting fluxes are then
compared with those predicted from the model atmospheres, integrated over the same bandpass.
We use the parallax measurements to constrain the surface gravity, if available. Otherwise, we assume a surface gravity of $\log$ g = 8.
The model atmospheres include the Ly $\alpha$ far red wing opacity \citep{kowalski06b} as well as non-ideal physics of
dense helium that includes refraction \citep{kowalski04}, ionization equilibrium \citep{kowalski07}, and the
non-ideal dissociation equilibrium of H$_2$ \citep{kowalski06a}. Using these models, \citet{kowalski06b} were able to reproduce the
ultraviolet to near-infrared spectral energy distributions of several cool white dwarfs including the DA star BPM 4729.
This star has a UV spectrum that can only be reproduced by pure hydrogen atmosphere models including the Ly $\alpha$ opacity.
This opacity occurs due to collisions of hydrogen atoms with H$_2$, and it dominates at $\lambda\leq6000$ \AA\ for cool
white dwarfs, i.e. when H$_2$ forms. Even though the Ly $\alpha$ opacity mainly supresses the flux in the ultraviolet, it causes
a significant redistribution of flux toward longer wavelengths.

We perform three separate sets of fits. The first set of fits involves pure hydrogen models with $T_{\rm eff}$ and $\log$ g as free
parameters. The second set of fits uses mixed H/He models and assumes that the
helium-to-hydrogen ratio is a free parameter as well.
Both sets of fits use models with $T_{\rm eff}\geq3500$ K and $\log$ g = $7-9$.
Finally, the last set of fits were performed with blackbody SEDs. \citet{kowalski06b} demonstrated that pure helium atmosphere
white dwarfs have SEDs similar to blackbodies, therefore if any of our targets have pure helium atmospheres, they should be
best explained by the third set of fits.

\section{Results}

Figure 1 displays the optical and infrared SEDs (error bars) for 44 cool white dwarfs
with $T_{\rm eff}<$ 6000 K.
Three of these stars (plotted at the end of the figure) are from the SDSS, and their photometry is in the SDSS $ugriz$
and 2MASS $JHK_s$ system. The remaining 41 stars have $BVRIJHK$ photometry available.
The expected fluxes from synthetic photometry of pure hydrogen and mixed H/He white dwarf model
atmospheres are shown as open and filled circles, respectively.
Since the pure hydrogen or mixed H/He models reproduce the SEDs better than the blackbody SEDs, 
the blackbody fits are not shown in any of the panels except for WD 2251$-$070.
This star is a DZ white dwarf.
Excluding the B-band flux which is affected by the presence of metals in the photosphere,
WD 2251$-$070 is the only object for which a blackbody SED is a better fit to the observed SED than pure hydrogen or mixed H/He models.

The differences between the pure hydrogen and mixed H/He
models are not statistically significant for the majority of our targets. For example, the predicted SEDs for the pure H and mixed H/He solutions for
WD 1656$-$062 are almost identical. One cannot differentiate between these two models based on the SED alone. WD 1656$-$062, like
19 other WDs in our sample, is a DA white dwarf. \citet{bergeron01} found that the pure
hydrogen models are in good agreement with H$\alpha$ spectroscopy of most stars in their (and our) sample.
Therefore, we adopt the pure hydrogen atmosphere solutions for all DAs.

A 5\% uncertainty in the $K-$band flux corresponds to a helium detection limit of He/H = 0.04, 0.36, and 2.14 for 4000, 5000, and
6000 K white dwarfs, respectively. Out of the 23 DC white dwarfs in our sample, 12 are best explained with mixed atmosphere models that have
He/H ratios above this detection limit. Hence,
we classify these white dwarfs as mixed H/He atmosphere objects. A few examples of the superior fits by mixed H/He fits include
WD 0029$-$032, WD 0747+073B, WD 1313$-$198, WD 1444$-$174, and SDSS J0753+4230. The differences between the best-fit pure hydrogen
and mixed H/He models are not significant for the remaining 11 DCs, and we assign pure hydrogen composition to them.
The physical parameters of the best fit atmosphere solutions as well as the spectral types of
our targets are presented in Table 2.

Only one star in Table 2 is assigned a pure helium composition. This result was expected
considering the work of \citet{kowalski06b}, which has satisfactorily demonstrated that the optical and near-infrared
colors of cool white dwarfs are explained fairly well with the new pure hydrogen models.
Yet this result is surprising from an evolutionary point of view,
since \citet{tremblay08} demonstrated that 45\% of white dwarfs in the range 7000 K $> T_{\rm eff} >$
5000 K have helium-rich atmospheres (with possible traces of hydrogen).
The mixed H/He composition stars make up 27\% of our sample. However, the helium to hydrogen ratio for these stars
is on the order of unity. Our result presents an important challenge to understanding the
spectral evolution of white dwarfs.

In any case, the choice of pure hydrogen or mixed H/He models do not have an important effect on understanding the mid-infrared
flux distributions of cool white dwarfs. Figure 1 demonstrates that the main difference between the best-fit pure hydrogen
and mixed H/He models is in the optical and near-infrared, and both sets of models agree in the mid-infrared.
Moreover, it shows that our grid of models reproduce the observed SEDs well in the entire range from 0.4$\mu$m to 8$\mu$m.
We do not find any significant mid-infrared flux deficits for these 44 cool white dwarfs. We now check to see how well
the models are able to reproduce the observations. \citet{tremblay07} found that the 12 cool white dwarfs that they analyzed
were on average fainter by 2\% in the 4.5 $\mu$m band, whereas they could fit the 8 $\mu$m photometry fairly well with their models.
\citet{debes07} and \citet{farihi08a} found that the models reproduce the observed mid-infrared fluxes within 5\%, slightly larger
than the IRAC calibration uncertainty of 3\% measured by \citet{reach05}.

Figure 2 presents the ratio of observed to predicted fluxes in all 11 filters from our analysis using the pure hydrogen
atmosphere fits. The IRAC 3.6, 4.5, 5.8, and 8 $\mu$m bands are labeled as I1, I2, I3, and I4, respectively.
Figure 3 shows the same ratios for the best-fit models to the SEDs (31 pure hydrogen and 12 mixed H/He atmosphere solutions given
in Table 2). A comparison of Figure 2 with Figure 3 shows that the use of mixed H/He models for the 12 DCs improves the agreement
between the observations and models.
Instead of looking at how well the models can reproduce the observations just in the mid-infrared, testing
the models in all filters gives us a better idea of how well the models work in general.
The photometric uncertainties in $VRI$ are 3\% and in $BJHK$ are 5\% \citep{bergeron01}.
Our photometric methods are consistent with those used for the IRAC
absolute calibration \citep{reach05}, and therefore the mid-infrared fluxes should be accurate to 3\%.

Table 3 presents the mean differences between the observed photometry and models for all filters. The pure hydrogen models
are consistent with the observations within 6\%. Excluding the IRAC 8$\mu$ band which has the lowest signal-to-noise ratio photometry,
the best fit models are consistent with observations within 5\%.
Given the 3-5\% uncertainty in the observations, 
we conclude that the optical, near-infrared, and IRAC photometry of cool white dwarfs are consistent with
white dwarf model atmospheres.

Subtle systematic trends as a function of effective temperature
are visible in Figure 2 and
3. Even though individual points in these trends
are not statistically off the one-to-one line, having a relatively large
sample of stars enables us to see these subtle effects. The $B-$ band and IRAC2
differences are perhaps the most convincing of these trends.  
Understanding these trends is useful for improving the current model atmosphere calculations.
The observed trend in the $B-$band is especially important as the $B-$band flux is mostly suppressed by the Ly $\alpha$ red wing
opacity in cool white dwarf atmospheres. Due to the uncertainty in the H$_2-$H potential energy surface
and transition dipole moments, the Ly $\alpha$ profiles used in \citet{kowalski06b} models
are reliable up to 6000 \AA. \citet{kowalski06b} extrapolated the Ly $\alpha$ profiles for longer wavelengths.
The Ly $\alpha$ absorption dominates up to 4500 \AA\ for $T_{\rm eff}=$ 5000 K
white dwarfs. However, it is important up to 6000 \AA\ and 7000 \AA\ for $T_{\rm eff}=$ 4500 K and 4000 K, respectively.
Therefore, assuming that the other absorption mechanisms are 
correctly described in the models, the observed drift with $T_{\rm eff}$
may indicate 
the need for extension of calculated Ly $\alpha$ profiles beyond $\lambda > 6000$ \AA.
However, we do not exclude other mechanisms that may cause the observed 
systematic trends.

Figure 4 shows four white dwarfs with {\em Spitzer} photometry that are perhaps affected by nearby bright stars.
Similar to the other 44 stars presented in previous figures, the SEDs for these stars are also consistent with model predictions,
and we do not find significant mid-infrared flux deficits.

The optical and infrared photometry and our IRS spectrum of the brightest target in our sample, the DA white dwarf Wolf 489,
are presented in Figure 5. 
The best fit model spectra with pure H and mixed H/He compositions are shown as solid and dashed lines, respectively.
The main difference between these two models are in the optical and they predict the same flux distribution in the mid-infrared.
The IRS spectrum is shown as a jagged line with error bars. Since Wolf 489 has a proper motion of $\approx3.9\arcsec$ yr$^{-1}$
\citep{lepine05}, the IRS centering was not perfect, and the measured fluxes from the spectrum were a factor of six
lower than the 
IRAC photometry. The centering problem resulted in lower signal-to-noise ratio data and low level fluctuations are visible in
the spectrum.
Since the IRS spectrum covers the IRAC 8$\mu$m band, we normalized it to match the 8$\mu$m photometry.
A comparion of the best fit models and the observations shows that the observed IRS spectrum is entirely consistent with
$T_{\rm eff}\approx$ 5000 K white dwarf models over the 5-15 $\mu$m range, further confirming our conclusion that the SEDs
of cool white dwarfs are consistent with model predictions.

\subsection{Ultracool White Dwarfs}

White dwarfs with $T_{\rm eff}$ thought to be below 4000 K are classified as ultracool. Ten years after the discovery of the
first ultracool white dwarf LHS 3250 \citep{harris99}, there are now more than a dozen known
\citep[see][and references therein]{gates04,harris08}. However, only a few of them have been studied
in detail using optical and near-infrared photometry and trigonometric parallax measurements \citep{bergeron02}.
The main difference between cool and ultracool white dwarfs is that several of these ultracool white dwarfs
show significant flux deficits in the near-infrared and even in the optical, which is most 
probably caused by the strong CIA absorption from H$_2$ in a dense, helium dominated atmosphere.
The SEDs of these white dwarfs cannot
be fit with any model atmospheres currently available. 

LHS 1126 is not classified as an ultracool white dwarf since its optical colors are consistent with $T_{\rm eff}=$ 5400 K
\citep{bergeron94}. However, it also displays significant flux deficits in the near- and mid-infrared. A remaining question is
whether all white dwarfs that display significant flux deficits in the near-infrared also display deficits in the mid-infrared or not.
The importance of this question is that
current CIA opacity calculations do not predict absorption in the mid-infrared. Therefore, finding several ultracool
white dwarfs with mid-infrared flux deficits would mean that there is a missing opacity source in current white dwarf atmosphere
models.

Three ultracool white dwarfs were included in our sample, LHS 3250, CE 51, and WD 0346+246. Unfortunately, WD 0346+246 was blended
with a brighter source in the IRAC images and we could not recover its photometry.
{\em Spitzer} photometry of CE 51 is also affected by a very bright nearby star.
\citet{ruiz01} assigned a temperature of 2730 K to CE 51 based on its optical photometry. Extending our models down to 2500 K, and
including the IRAC photometry in our fits does not change the best-fit temperature much.
We find a best-fit solution of 2660 K for a pure hydrogen composition. However,
the best-fit model is not a good representation of the optical and mid-infrared photometry.
Since our IRAC photometry is questionable for this star, we cannot draw any other conclusions at this time.

LHS 3250 was close to an extended source in our IRAC images, however we are able to recover the photometry using a 2 pixel aperture.
Figure 6 shows LHS 3250 images from 1988 to 2007. Our {\em Spitzer} IRAC images were taken in August 2006, 9 months before the
$I-$band image from the MDM 2.4m telescope was taken. LHS 3250 has a proper motion of
($\mu_{\alpha}\ cos \delta, \mu_{\delta}) = (-0.541$, +0.148 mas yr$^{-1}$; Munn et al. 2004) ,
therefore the position measured from the $I-$band images should be essentially the same as the position in our
IRAC images. This position is marked with open circles in the MDM and IRAC images; LHS 3250 is clearly detected.

LHS 3250 is quite faint in the mid-infrared. From the BCD frames, we measure average fluxes of 7.0 $\pm$ 1.7, 5.8 $\pm$ 2.3, 8.1
$\pm$ 4.9, and 13 $\pm$ 6.4 $\mu$Jy in the 3.6, 4.5, 5.8, and 8 $\mu$m bands, respectively. Using the combined mosaic images,
we measure 7.0, 6.0, 6.1, and 10.7 $\mu$Jy in the same bands, respectively. Photometry from the BCD frames and the mosaic frames
are consistent within the errors.
The LHS 3250 SED compared to the best fit mixed H/He atmosphere model is shown in the top panel in Figure 7. 
The effect of CIA opacity is clearly seen in the models; there is a big dip
in the predicted fluxes up to about 3 $\mu$m due to H$_2$ 1--0 dipole absorption centered at 2.4 $\mu$m.
The models obviously fail to reproduce the observations in all bands.
As in previous attempts by other authors \citep{harris99,bergeron02}, the SED cannot be explained with current white dwarf
model atmospheres.

The bottom panel in Figure 7 compares the LHS 3250 SED with LHS 1126.
We normalized the LHS 1126 SED to match LHS 3250 in the $B-$band. Since LHS 1126 is more than an order of magnitude
brighter, its infrared fluxes are more accurate. The infrared portion of the LHS 1126 SED is best fit with a power law index of
$-$1.99. On the other hand, the infrared portion of the LHS 3250 SED, excluding the 5.8 and 8 $\mu$m photometry due to large error bars, fits a power law
index of $-$2.07. However, the power law fit is not as good a fit as for LHS 1126. The fluxes measured between 3.6 and 8 $\mu$m
are consistent with each other within the errors.

\section{Discussion}

\subsection{Cool White Dwarf Atmospheres in the Infrared}

We now have extended the earlier work on {\em Spitzer} observations of cool white dwarfs \citep{kilic06a,kilic08a} down to 4000 K
using a well studied sample of nearby white dwarfs. Our fits to the $BVRIJHK$ and $3.6-8 \mu$m photometry of $\approx$50 cool
white dwarfs show that current state of the art white dwarf model atmospheres are able to reproduce the observations within
5\%. The only exception to this is LHS 1126, which shows significant flux deficits in the mid-infrared compared to the models.
In addition, our models are not able to explain the observed optical and infrared SED of the ultracool white dwarf LHS 3250.
More theoretical work on understanding the properties 
of dense helium-rich, mixed H/He white dwarf atmospheres is probably needed 
to understand the spectra of stars like this.
A comparison of the SEDs for LHS 1126 and LHS 3250 shows that a power law may explain the infrared flux distributions of both
stars, however, the
LHS 3250 SED shows evidence of a flat or increasing mid-infrared flux distribution, and therefore its 
SED remains puzzling. Obtaining higher quality mid-infrared data for LHS 3250 and extending the wavelength coverage further to
the red will be important in understanding
LHS 3250 and other ultracool white dwarf atmospheres. This will be possible with the
Mid-Infrared Instrument (MIRI) on the James Webb Space Telescope. 

\subsection{Infrared Excess from Dust Disks}

Previous ground- and space-based studies of metal-rich white dwarfs showed that $\approx$20\% of hydrogen atmosphere
white dwarfs with $T_{\rm eff}\geq$ 7400 K display infrared excesses from dust disks \citep[see][and references therein]{kilic08b}.
However, the majority of these disks are found around stars hotter than 10000 K and younger than 1 Gyr.
The fraction of dusty white dwarfs goes down to 1-3\% for previous {\em Spitzer} observations of $\approx$200 white dwarfs with somewhat
less metallicity bias \citep{farihi08b}. \citet{mullally07} found 2 dusty stars out of 124 white dwarfs, corresponding to a fraction of
$1.6^{+2.1}_{-0.5}$ \%.
Assuming optically thick and flat disk models, these disks are within 
1$R_\odot$ of white dwarfs \citep{jura07,vonhippel07}. An important step in understanding the formation and evolution of these disks
is the disk lifetimes and whether the dust can survive for the lifetime of the white dwarf or not.

Previous {\em Spitzer} observations of white dwarfs targeted mostly stars with $T_{\rm eff}>$ 6000 K and ages younger than 2 Gyr.
Extending the temperature limit down to 4000 K corresponds to extending the ages up to 9 Gyr for typical 0.6 $M_\odot$ white
dwarfs \citep{bergeron95}.
Excluding the ultracool white dwarfs from our sample, none of the 48 cool white dwarfs (Figure 1 and 4) in our sample
shows excess infrared radiation from disks. However, cooler stellar effective temperatures would mean that if disks are present,
they will be substantially cooler than the disks around warmer white dwarfs.
For example, if we take the disk around G29-38 as a prototype and put it around a 5000 K white dwarf, the disk would have
an inner temperature of $\approx$ 600 K and an outer temperature of 350 K.

Figure 8 shows several flat disk models \citep{jura03} for a 5000 K white dwarf at 20 pc. The solid line shows the total flux expected
from the photosphere of a 5000 K white dwarf and a disk with an inclination angle 45$^o$ and the same inner-outer radii as G29-38.
The short-dashed line
shows the flux expected from a disk further away from the star, with parameters similar to G166$-$58 \citep{farihi08a}.
The other dashed and dashed-dotted lines show the differences in disk emission depending on the inner radii of the disks. Here we only
explore the disks out to 1.2 $R_\odot$ as it corresponds to the Roche limit for typical white dwarfs \citep{vonhippel07,farihi08a}.
Adopting an accuracy of 5\% in the Spitzer IRAC bands, the solid horizontal line shows 3$\sigma$ excess above the photospheric emission
at 8$\mu$m. It is evident from this figure that, if disks survive for several billion years or if they are replenished, they should be detected
in our IRAC observations if the inner radii of the disks are smaller than 0.5 $R_\odot$.
Colder disks with larger inner radii would be visible at longer wavelengths.

Non-detection of disks around our 48 targets means that the frequency of warm disks around cool white dwarfs is $0.8^{+1.5}_{-0.8}$ \%,
still consistent with the frequency of disks for warmer white dwarfs. 

\acknowledgements
MK thanks D. Winget and S. Kenyon for useful discussions.
Support for this work was provided by NASA through the Spitzer Space Telescope Fellowship Program, under an award from Caltech, and
also award projects NBR 1359963 and 1288635 issued by JPL/Caltech to Ohio State University and the University of Texas.
PMK thanks D. Marx and the theoretical chemistry group at 
Ruhr-Universit\"at Bochum for their hospitality and support.
This work is based in part on observations made with the {\em Spitzer Space Telescope}, which is operated by the Jet Propulsion
Laboratory, California Institute of Technology under NASA contract 1407.

\clearpage
\begin{deluxetable}{llcccc}
\tabletypesize{\footnotesize}
\tablecolumns{6}
\tablewidth{0pt}
\tablecaption{IRAC Photometry of Cool White Dwarfs}
\tablehead{
\colhead{Object}&
\colhead{Name}&
\colhead{3.6$\mu$m}&
\colhead{4.5$\mu$m}&
\colhead{5.8$\mu$m}&
\colhead{8.0$\mu$m}
\\
        &      & (mJy) & (mJy) & (mJy) & (mJy)}
\startdata
WD 0011$-$399 & J0014$-$3937 & 0.1186 $\pm$ 0.0039 & 0.0798 $\pm$ 0.0029 & 0.0597 $\pm$ 0.0073 & 0.0330 $\pm$ 0.0132 \\
WD 0029$-$032 & LHS1093 & 0.2591 $\pm$ 0.0085 & 0.1824 $\pm$ 0.0067 & 0.1274 $\pm$ 0.0264 & 0.0502 $\pm$ 0.0193 \\
WD 0042$-$337 & F351$-$50 & 0.0613 $\pm$ 0.0022 & 0.0426 $\pm$ 0.0018 & 0.0283 $\pm$ 0.0120 & 0.0216 $\pm$ 0.0039 \\
WD 0048$-$207 & LHS1158 & 0.2308 $\pm$ 0.0081 & 0.1550 $\pm$ 0.0057 & 0.0965 $\pm$ 0.0088 & 0.0453 $\pm$ 0.0018 \\
WD 0102+210.2 & LHS5024 & 0.1129 $\pm$ 0.0036 & 0.0779 $\pm$ 0.0028 & 0.0505 $\pm$ 0.0103 & 0.0249 $\pm$ 0.0241 \\
WD 0102+210.1 & LHS5023 & 0.1018 $\pm$ 0.0033 & 0.0715 $\pm$ 0.0026 & 0.0536 $\pm$ 0.0104 & 0.0224 $\pm$ 0.0161 \\
WD 0112$-$018 & LHS1219 & 0.1383 $\pm$ 0.0044 & 0.0908 $\pm$ 0.0032 & 0.0598 $\pm$ 0.0157 & 0.0324 $\pm$ 0.0144 \\
WD 0117$-$145 & LHS1233 & 0.3033 $\pm$ 0.0098 & 0.2032 $\pm$ 0.0071 & 0.1239 $\pm$ 0.0223 & 0.1083 $\pm$ 0.0266 \\
WD 0121+401 & G133$-$8 & 0.2216 $\pm$ 0.0074 & 0.1465 $\pm$ 0.0055 & 0.0831 $\pm$ 0.0098 & 0.0567 $\pm$ 0.0102 \\
WD 0133$-$548 & WD0135$-$546 & 0.1044 $\pm$ 0.0034 & 0.0686 $\pm$ 0.0024 & 0.0457 $\pm$ 0.0050 & 0.0263 $\pm$ 0.0049 \\
WD 0202$-$055 & WD0205$-$053 & 0.1057 $\pm$ 0.0035 & 0.0732 $\pm$ 0.0033 & 0.0399 $\pm$ 0.0078 & 0.0277 $\pm$ 0.0066 \\
WD 0222+648$^a$ & LHS1405 & 0.1358 $\pm$ 0.0352 & 0.0951 $\pm$ 0.0244 & 0.0653 $\pm$ 0.0159 & 0.0401 $\pm$ 0.0100 \\
WD 0230$-$144 & LHS1415 & 0.7045 $\pm$ 0.0220 & 0.4654 $\pm$ 0.0152 & 0.3185 $\pm$ 0.0164 & 0.1754 $\pm$ 0.0180 \\
WD 0357+081 & LHS1617 & 0.6134 $\pm$ 0.0193 & 0.4106 $\pm$ 0.0136 & 0.2526 $\pm$ 0.0159 & 0.1416 $\pm$ 0.0140 \\
WD 0407+197 & LHS1636 & 0.1354 $\pm$ 0.0050 & 0.0815 $\pm$ 0.0040 & 0.0504 $\pm$ 0.0244 & 0.0376 $\pm$ 0.0289 \\
WD 0423+044 & LHS1670 & 0.2655 $\pm$ 0.0088 & 0.1789 $\pm$ 0.0069 & 0.1185 $\pm$ 0.0123 & 0.0785 $\pm$ 0.0189 \\
WD 0503$-$174 & LHS1734 & 0.5840 $\pm$ 0.0187 & 0.3797 $\pm$ 0.0125 & 0.2536 $\pm$ 0.0164 & 0.1455 $\pm$ 0.0142 \\
WD 0551+468$^a$ & LHS1801 & 0.2527 $\pm$ 0.0089 & 0.1608 $\pm$ 0.0068 & 0.0979 $\pm$ 0.0134 & 0.0510 $\pm$ 0.0205 \\
WD 0657+320 & LHS1889 & 0.4238 $\pm$ 0.0135 & 0.2861 $\pm$ 0.0097 & 0.1969 $\pm$ 0.0140 & 0.1279 $\pm$ 0.0169 \\
WD 0727+482AB$^b$ & G107$-$70 & 2.5598 $\pm$ 0.0782 & 1.7041 $\pm$ 0.0525 & 1.1482 $\pm$ 0.0395 & 0.6676 $\pm$ 0.0256 \\
WD 0743$-$340 & vB3 & \nodata $\pm$ \nodata & 0.3585 $\pm$ 0.0125 & \nodata $\pm$ \nodata  & 0.1044 $\pm$ 0.0080 \\
WD 0747+073A & LHS239 & 0.3987 $\pm$ 0.0127 & 0.2739 $\pm$ 0.0094 & 0.1668 $\pm$ 0.0141 & 0.1017 $\pm$ 0.0202 \\
WD 0747+073B & LHS240 & 0.4552 $\pm$ 0.0144 & 0.3085 $\pm$ 0.0103 & 0.1990 $\pm$ 0.0168 & 0.1048 $\pm$ 0.0155 \\
SDSS J0753+4230 & J0753+4230 & 0.2028 $\pm$ 0.0068 & 0.1427 $\pm$ 0.0056 & 0.0963 $\pm$ 0.0093 & 0.0565 $\pm$ 0.0189 \\
WD 0851$-$246$^a$ & CE51 & 0.1010 $\pm$ 0.0066 & 0.0790 $\pm$ 0.0064 & 0.0677 $\pm$ 0.0164 & \nodata $\pm$ \nodata \\
WD 1022+009 & LHS282 & 0.0875 $\pm$ 0.0029 & 0.0557 $\pm$ 0.0023 & 0.0274 $\pm$ 0.0093 & 0.0183 $\pm$ 0.0126 \\
WD 1108+207 & LHS2364 & 0.1892 $\pm$ 0.0067 & 0.1197 $\pm$ 0.0055 & 0.0955 $\pm$ 0.0133 & 0.0310 $\pm$ 0.0346 \\
WD 1136$-$286 & ESO439$-$26 & 0.0144 $\pm$ 0.0010 & 0.0077 $\pm$ 0.0014 & 0.0186 $\pm$ 0.0110 & 0.0031 $\pm$ 0.0003 \\
WD 1153+135 & LHS2478 & 0.1631 $\pm$ 0.0056 & 0.1050 $\pm$ 0.0046 & 0.0800 $\pm$ 0.0200 & 0.0559 $\pm$ 0.0188 \\
WD 1247+550$^a$ & LP131$-$66 & 0.2353 $\pm$ 0.0082 & 0.1569 $\pm$ 0.0060 & 0.1025 $\pm$ 0.0111 & 0.0648 $\pm$ 0.0106 \\
WD 1257+037 & LHS2661 & 0.5908 $\pm$ 0.0187 & 0.3859 $\pm$ 0.0126 & 0.2668 $\pm$ 0.0163 & 0.1340 $\pm$ 0.0161 \\
WD 1300+263 & LHS2673 & 0.0749 $\pm$ 0.0026 & 0.0517 $\pm$ 0.0023 & 0.0399 $\pm$ 0.0061 & 0.0121 $\pm$ 0.0103 \\
WD 1313$-$198 & LHS2710 & 0.2037 $\pm$ 0.0069 & 0.1336 $\pm$ 0.0052 & 0.1003 $\pm$ 0.0147 & 0.0722 $\pm$ 0.0240 \\
SDSS J1313+0226 & J1313+0226 & 0.1327 $\pm$ 0.0044 & 0.0901 $\pm$ 0.0031 & 0.0615 $\pm$ 0.0069 & 0.0379 $\pm$ 0.0128 \\
WD 1334+039 & Wolf489 & 2.6091 $\pm$ 0.0805 & 1.7405 $\pm$ 0.0548 & 1.1851 $\pm$ 0.0477 & 0.6677 $\pm$ 0.0329 \\
WD 1345+238 & LP380$-$5 & 1.1767 $\pm$ 0.0363 & 0.7908 $\pm$ 0.0248 & 0.5380 $\pm$ 0.0226 & 0.3165 $\pm$ 0.0185 \\
WD 1346+121 & LHS2808 & 0.1001 $\pm$ 0.0035 & 0.0654 $\pm$ 0.0043 & 0.0411 $\pm$ 0.0125 & 0.0330 $\pm$ 0.0036 \\
WD 1444$-$174 & LHS378 & 0.4579 $\pm$ 0.0146 & 0.3142 $\pm$ 0.0106 & 0.2212 $\pm$ 0.0174 & 0.1589 $\pm$ 0.0162 \\
WD 1602+010 & LHS3151 & 0.1701 $\pm$ 0.0058 & 0.1151 $\pm$ 0.0047 & 0.0671 $\pm$ 0.0114 & 0.0498 $\pm$ 0.0164 \\
WD 1653+630 & LHS3250 & 0.0070 $\pm$ 0.0017 & 0.0058 $\pm$ 0.0023 & 0.0081 $\pm$ 0.0049 & 0.0130 $\pm$ 0.0064 \\
WD 1656$-$062 & LP686$-$32 & 0.1826 $\pm$ 0.0069 & 0.1203 $\pm$ 0.0055 & 0.0746 $\pm$ 0.0186 & 0.0203 $\pm$ 0.0156 \\
WD 1820+609 & G227$-$28 & 1.0795 $\pm$ 0.0333 & 0.7283 $\pm$ 0.0229 & 0.4773 $\pm$ 0.0198 & 0.2700 $\pm$ 0.0201 \\
WD 2002$-$110 & LHS483 & 0.3407 $\pm$ 0.0115 & 0.2390 $\pm$ 0.0090 & 0.1359 $\pm$ 0.0157 & 0.0956 $\pm$ 0.0205 \\
WD 2048+263 & G187$-$8 & 0.9917 $\pm$ 0.0307 & 0.6574 $\pm$ 0.0208 & 0.4573 $\pm$ 0.0206 & 0.2591 $\pm$ 0.0146 \\
WD 2054$-$050$^a$ & vB11 & 0.5335 $\pm$ 0.0235 & 0.3487 $\pm$ 0.0270 & 0.1060 $\pm$ 0.0820 & 0.0778 $\pm$ 0.0266 \\
SDSS J2116$-$0724 & J2116$-$0724 & 0.1442 $\pm$ 0.0051 & 0.0991 $\pm$ 0.0042 & 0.0735 $\pm$ 0.0327 & 0.0532 $\pm$ 0.0214 \\
WD 2248+293 & G128$-$7 & 0.7773 $\pm$ 0.0242 & 0.5090 $\pm$ 0.0165 & 0.3545 $\pm$ 0.0184 & 0.1781 $\pm$ 0.0142 \\
WD 2251$-$070 & LP701$-$29 & 1.3402 $\pm$ 0.0412 & 0.9103 $\pm$ 0.0287 & 0.6108 $\pm$ 0.0273 & 0.3791 $\pm$ 0.0252 \\
WD 2316$-$064 & LHS542 & 0.1237 $\pm$ 0.0040 & 0.0799 $\pm$ 0.0030 & 0.0596 $\pm$ 0.0085 & 0.0283 $\pm$ 0.0165 \\
WD 2343$-$481 & WD2346$-$478 & 0.1225 $\pm$ 0.0039 & 0.0752 $\pm$ 0.0028 & 0.0452 $\pm$ 0.0052 & 0.0323 $\pm$ 0.0116 \\
WD 2345$-$447 & ESO292$-$43 & 0.0948 $\pm$ 0.0031 & 0.0627 $\pm$ 0.0023 & 0.0431 $\pm$ 0.0067 & 0.0319 $\pm$ 0.0121 \\
\enddata
\tablecomments{(a)- The photometry is likely to be affected by nearby bright star. (b)- WD 0727+482A and WD 0727+482B are unresolved
in IRAC images. The photometry given here is the combined flux from both stars.}
\end{deluxetable}

\clearpage
\begin{deluxetable}{lccccc}
\tabletypesize{\footnotesize}
\tablecolumns{5}
\tablewidth{0pt}
\tablecaption{Atmospheric Parameters of Cool White Dwarfs}
\tablehead{
\colhead{Object}&
\colhead{Spectral}&
\colhead{Comp}&
\colhead{$T_{\rm eff}$}&
\colhead{$\log$ g}&
\colhead{He/H}\\
        & Type &  & (K) & & }
\startdata
WD 0011$-$399 & DC  & H    & 4700 &  8   & \nodata \\
WD 0029$-$032 & DC  & He/H & 4590 & 7.96 & 0.3 \\
WD 0042$-$337 & DC  & H    & 4040 &  8   & \nodata \\
WD 0048$-$207 & DA  & H    & 5190 &  8   & \nodata \\
WD 0102+210.2 & DC  & H    & 4910 &  8   & \nodata \\
WD 0102+210.1 & DA  & H    & 5320 &  8   & \nodata \\
WD 0112$-$018 & DA  & H    & 5410 &  8   & \nodata \\
WD 0117$-$145 & DA  & H    & 5070 & 7.75 & \nodata \\
WD 0121+401   & DA  & H    & 5330 & 7.94 & \nodata \\
WD 0133$-$548 & DC  & He/H & 4660 &  8   & 0.2  \\
WD 0202$-$055 & DC  & H    & 4150 &  8   & \nodata \\
WD 0230$-$144 & DA  & H    & 5470 &  8.15 & \nodata \\
WD 0357+081   & DA  & H    & 5490 &  8.07 & \nodata \\
WD 0407+197   & DC  & H    & 5250 &  8  & \nodata \\
WD 0423+044   & DA  & H    & 4950 &  8  & \nodata \\
WD 0503$-$174 & DA  & H    & 5410 &  7.72  & \nodata \\
WD 0657+320   & DA  & H    & 4930 &  8.09  & \nodata \\
WD 0743$-$340 & DC  & H    & 4630 &  8.15  & \nodata \\
WD 0747+073A  & DA  & H    & 4500 &  7.97  & \nodata \\
WD 0747+073B  & DC  & He/H & 4700 &  7.97 & 0.8 \\
SDSS J0753+4230& DC & He/H & 4540 &  8 & 0.7  \\
WD 1022+009   & DA  & H    & 5400 &  7.65  & \nodata \\
WD 1108+207   & DC  & H    & 4760 &  8.18  & \nodata  \\
WD 1136$-$286 & DC  & H    & 4630 &  9.12  & \nodata \\
WD 1153+135   & DC  & He/H & 4830 &  8 & 0.3  \\
WD 1257+037   & DA  & H    & 5640 &  8.22  & \nodata \\
WD 1300+263   & DC  & H    & 4360 &  8.16  & \nodata \\
WD 1313$-$198 & DC  & He/H & 5110 &  8.24 & 1.7 \\
SDSS J1313+0226& DC & H    & 4230 &  8  & \nodata \\
WD 1334+039   & DA  & H    & 5030 &  8.01  & \nodata \\
WD 1345+238   & DA  & H    & 4690 &  7.91  & \nodata \\
WD 1346+121   & DC  & He/H & 4710 &  8 & 0.9  \\
WD 1444$-$174 & DC  & He/H & 4820 &  8.31 & 0.7 \\
WD 1602+010   & DC  & H    & 4880 &  8  & \nodata \\
WD 1656$-$062 & DA  & H    & 5550 &  8.09  & \nodata \\
WD 1820+609   & DA  & H    & 4900 &  7.97  & \nodata  \\
WD 2002$-$110 & DC  & He/H & 4640 &  8.23 & 0.3  \\
WD 2048+263   & DA  & H    & 5110 &  7.65  & \nodata \\
SDSS J2116$-$0724& DC&He/H & 4650 &  8 & 0.3  \\
WD 2248+293   & DA  & H    & 5650 &  7.65  & \nodata \\
WD 2251$-$070 & DZ  & He   & 4790 & Blackbody \\
WD 2316$-$064 & DC  & H    & 4700 &  8.18  & \nodata \\
WD 2343$-$481 & DA  & H    & 5130 &  8  & \nodata \\
WD 2345$-$447 & DC  & He/H & 5330 &  8.70 & 2.0  \\
\enddata
\end{deluxetable}

\clearpage
\begin{deluxetable}{lcccc}
\tablecolumns{5}
\tablewidth{0pt}
\tablecaption{Mean \% Differences Between the Observed Photometry and Models}
\tablehead{
\colhead{}&
\colhead{pure H}&
\colhead{pure H}&
\colhead{Best-fit}&
\colhead{Best-fit}\\
Filter  & Mean & Wt-Mean & Mean & Wt-Mean}
\startdata
B  & $-5.7 \pm 3.7$ & $-6.0 \pm 0.7$ & $-4.9 \pm 3.7$ & $-5.2 \pm 0.7$ \\
V  & $+1.5 \pm 2.7$ & $+1.3 \pm 0.4$ & $+0.9 \pm 2.2$ & $+0.8 \pm 0.4$ \\
R  & $+3.8 \pm 3.2$ & $+3.6 \pm 0.5$ & $+2.8 \pm 2.8$ & $+2.7 \pm 0.5$ \\
I  & $+2.9 \pm 4.3$ & $+2.5 \pm 0.5$ & $+1.6 \pm 3.6$ & $+1.4 \pm 0.4$ \\
J  & $+6.0 \pm 4.4$ & $+5.7 \pm 0.8$ & $+5.0 \pm 5.3$ & $+4.5 \pm 0.8$ \\
H  & $-0.9 \pm 6.5$ & $-1.8 \pm 0.8$ & $+0.7 \pm 5.0$ & $+0.2 \pm 0.8$ \\
K  & $-0.9 \pm 6.6$ & $-1.9 \pm 0.8$ & $+1.8 \pm 5.1$ & $+1.4 \pm 0.8$ \\
IRAC1  & $-2.0 \pm 4.2$ & $-2.5 \pm 0.5$ & $-1.1 \pm 3.6$ & $-1.5 \pm 0.5$ \\
IRAC2  & $-5.0 \pm 4.6$ & $-4.6 \pm 0.5$ & $-4.5 \pm 4.4$ & $-4.2 \pm 0.5$ \\
IRAC3  & $-0.1 \pm 30.6$ & $-3.1 \pm 1.3$ & $-0.1 \pm 30.6$ & $-3.0 \pm 1.3$ \\
IRAC4  & $-5.7 \pm 21.1$ & $-11.4 \pm 1.6$ & $-6.3 \pm 20.3$ & $-11.6 \pm 1.6$ \\
\enddata
\end{deluxetable}

\clearpage
\begin{figure}
\hspace{-0.8in}
\includegraphics[angle=-90,scale=.75]{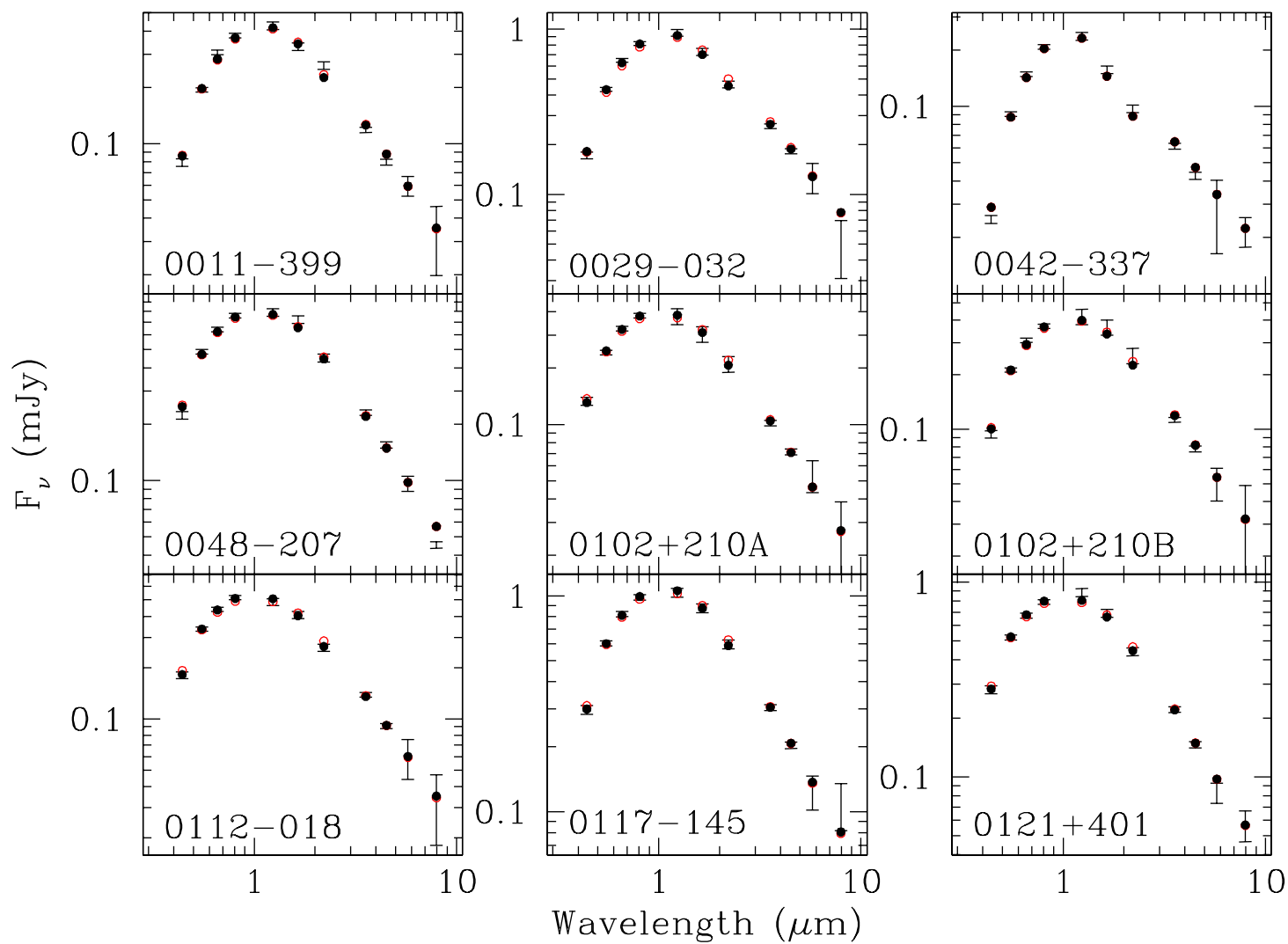}
\caption{Spectral energy distributions of cool white dwarfs observed in our program.
The observed fluxes are shown as error bars, whereas the expected flux distributions
from pure hydrogen and mixed H/He atmosphere models are shown as open and filled circles,
respectively. Instead of the mixed H/He atmosphere models, only the panel for WD 2251$-$070
shows the predicted flux distribution for a blackbody SED as filled circles.}
\end{figure}

\clearpage
\begin{figure}
\hspace{-0.8in}
\includegraphics[angle=-90,scale=.75]{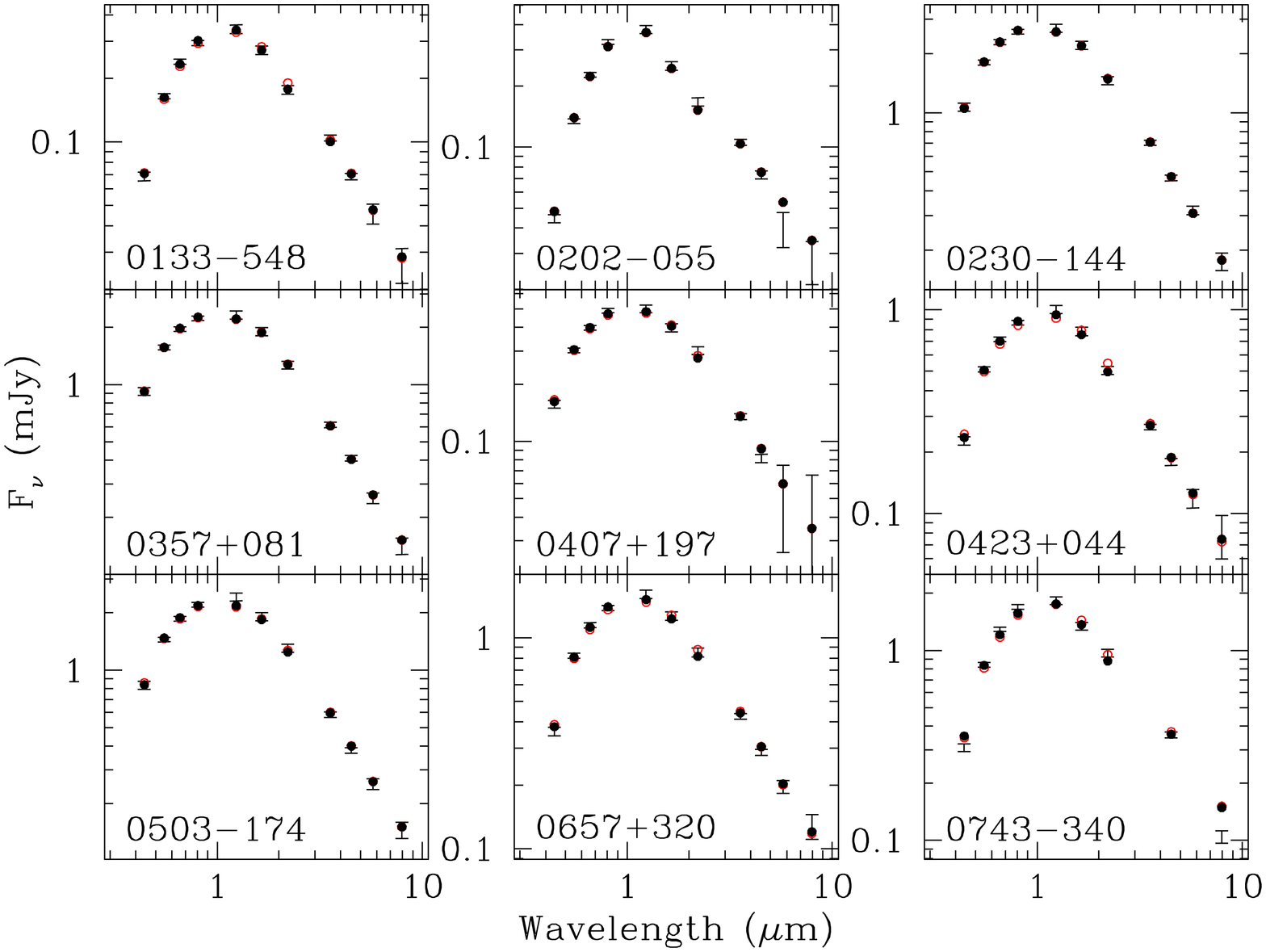}
\figurenum{1}
\caption{contd.}
\end{figure}

\clearpage
\begin{figure}
\hspace{-0.8in}
\includegraphics[angle=-90,scale=.75]{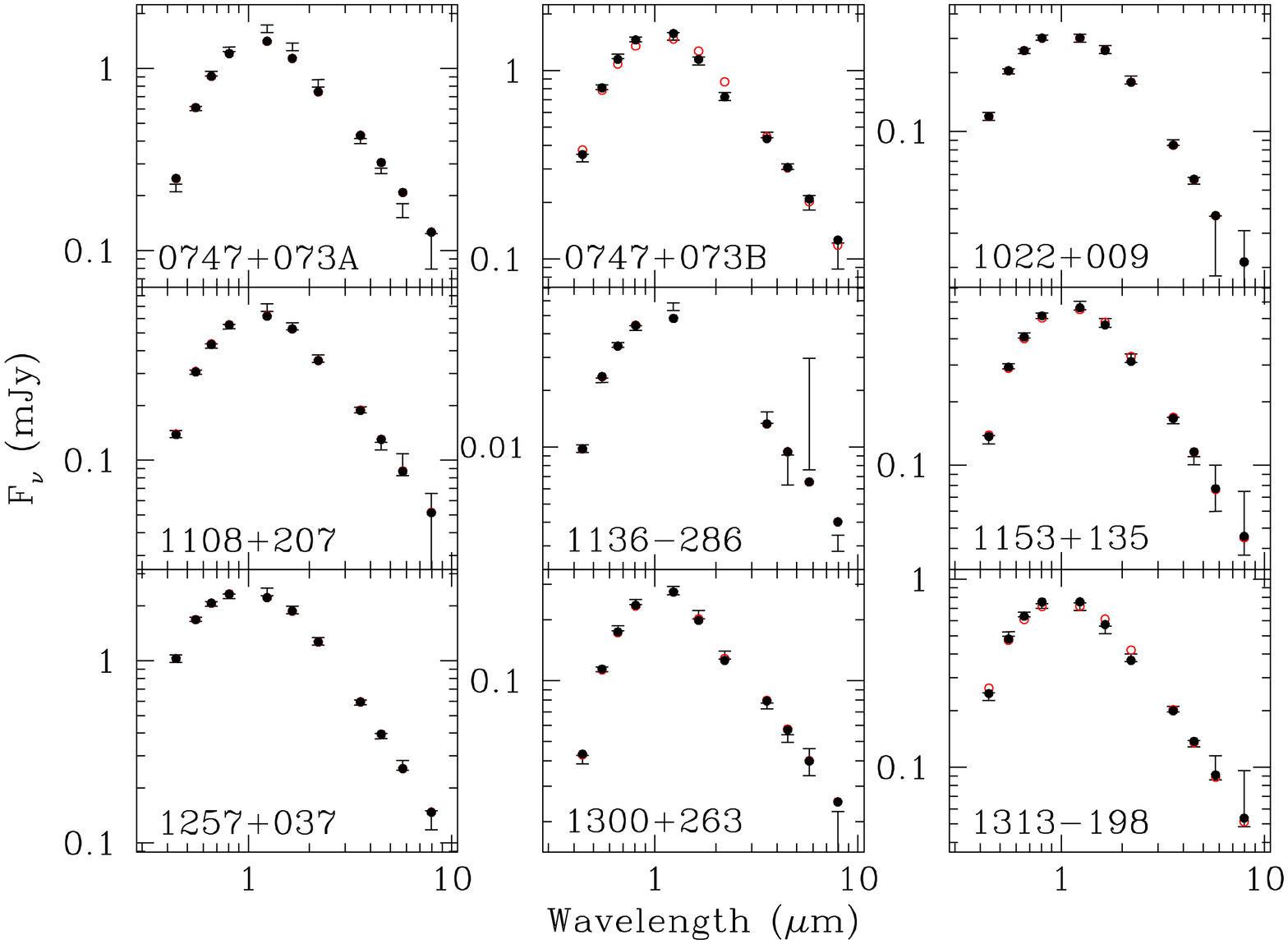}
\figurenum{1}
\caption{contd.}
\end{figure}

\clearpage
\begin{figure}
\hspace{-0.8in}
\includegraphics[angle=-90,scale=.75]{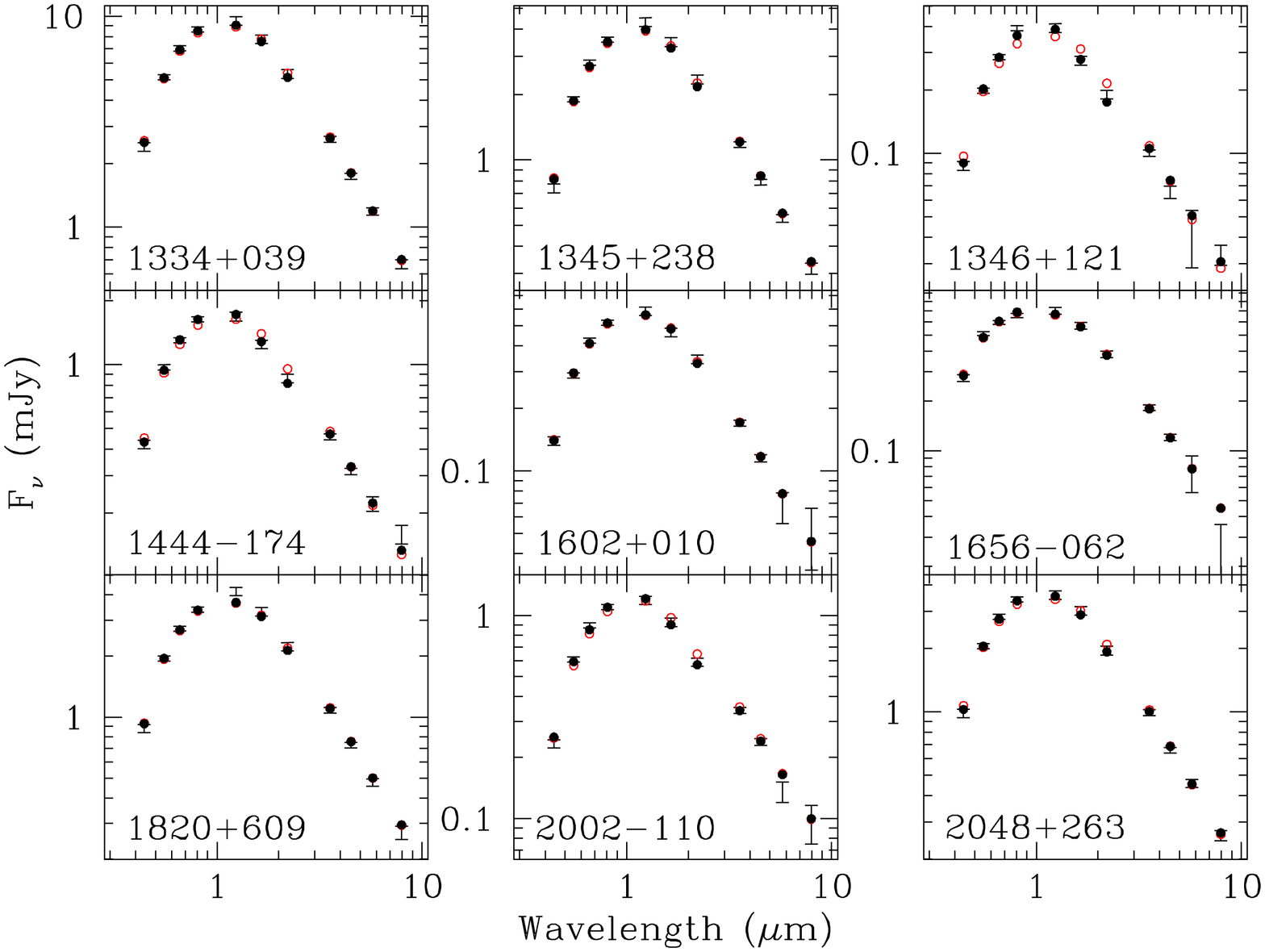}
\figurenum{1}
\caption{contd.}
\end{figure}

\clearpage
\begin{figure}
\hspace{-0.8in}
\includegraphics[angle=-90,scale=.75]{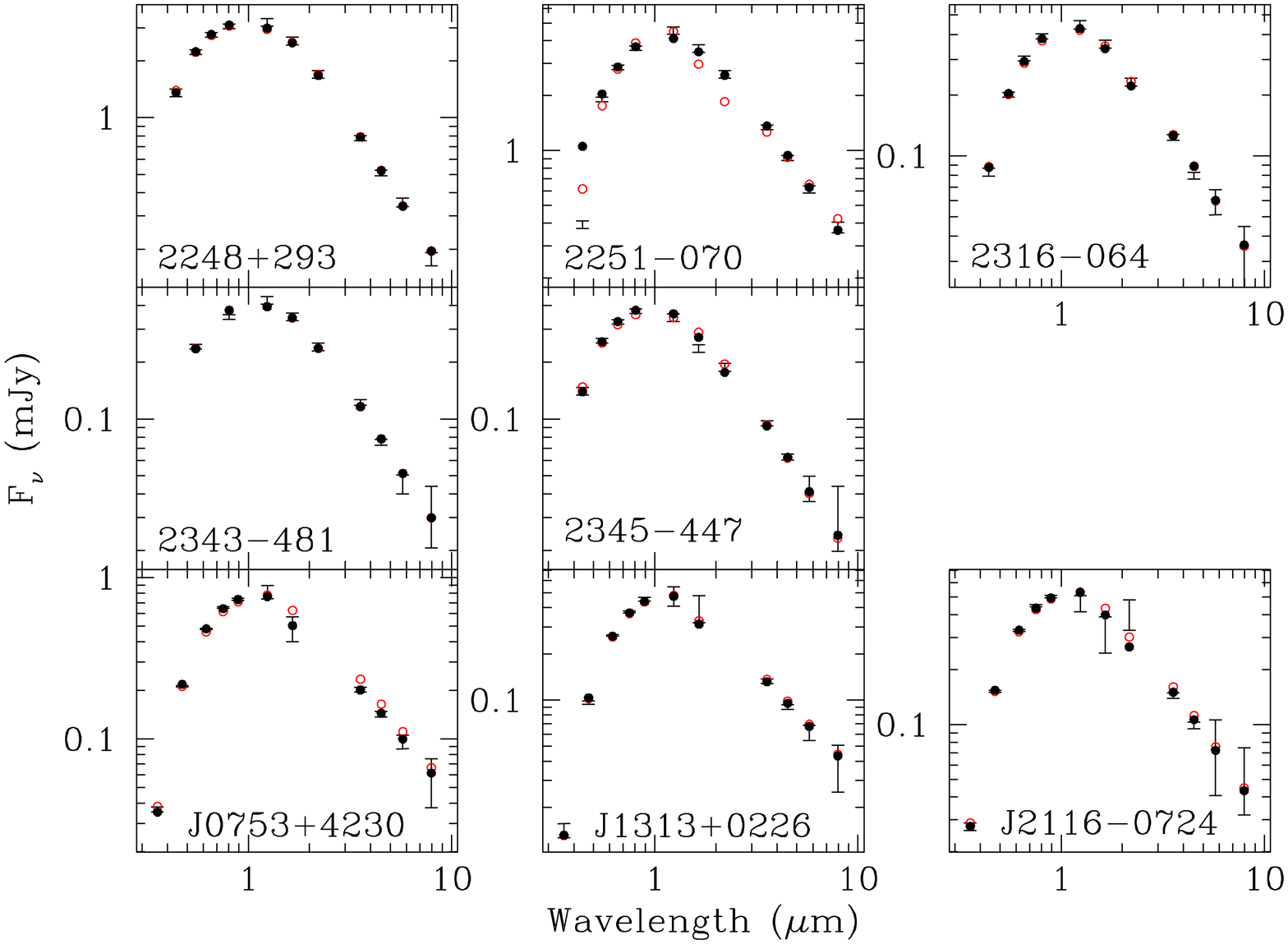}
\figurenum{1}
\caption{contd.}
\end{figure}

\clearpage
\begin{figure}
\plotone{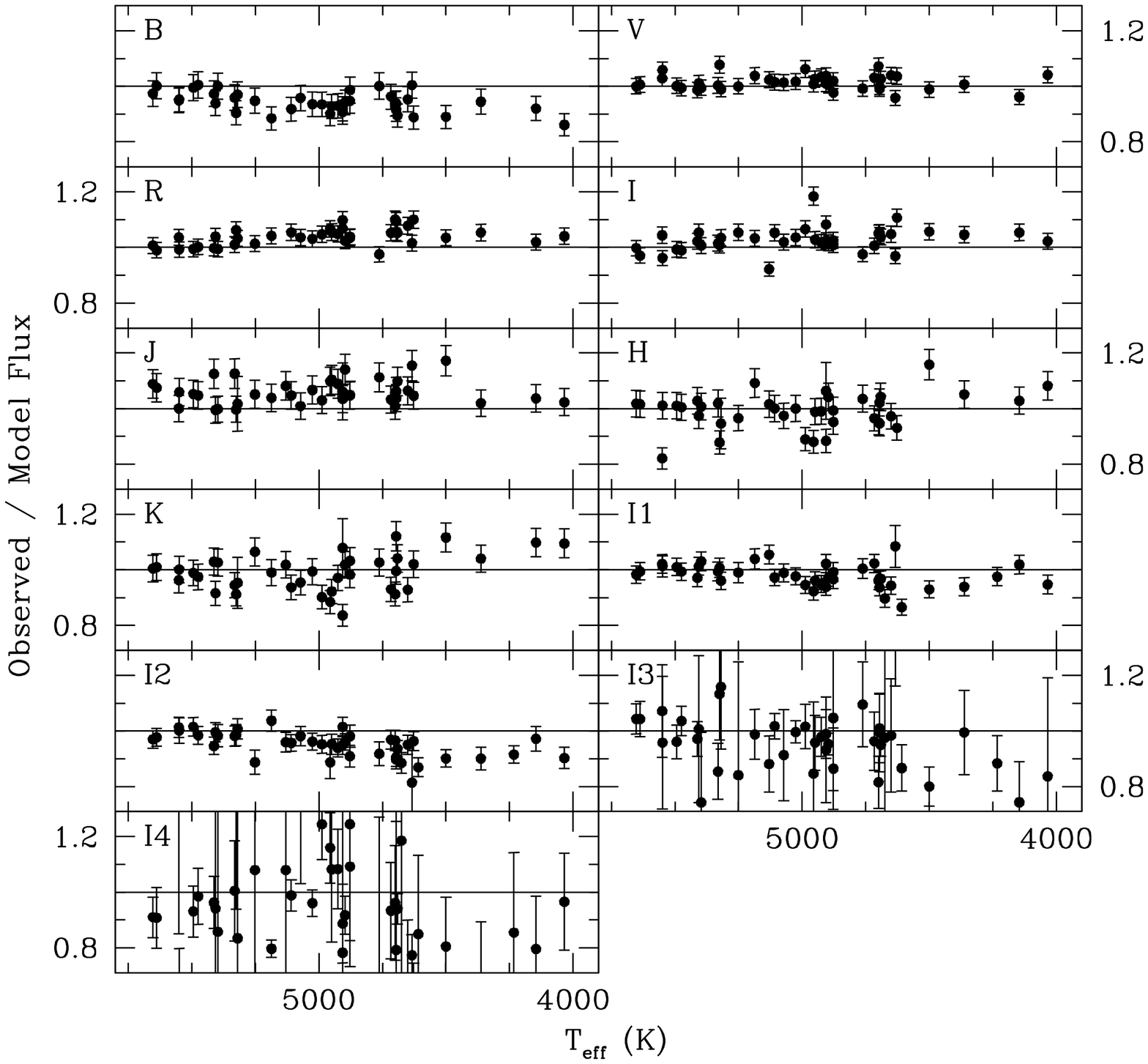}
\caption{Ratio of the observed vs. model fluxes for the pure hydrogen models.}
\end{figure}

\clearpage
\begin{figure}
\plotone{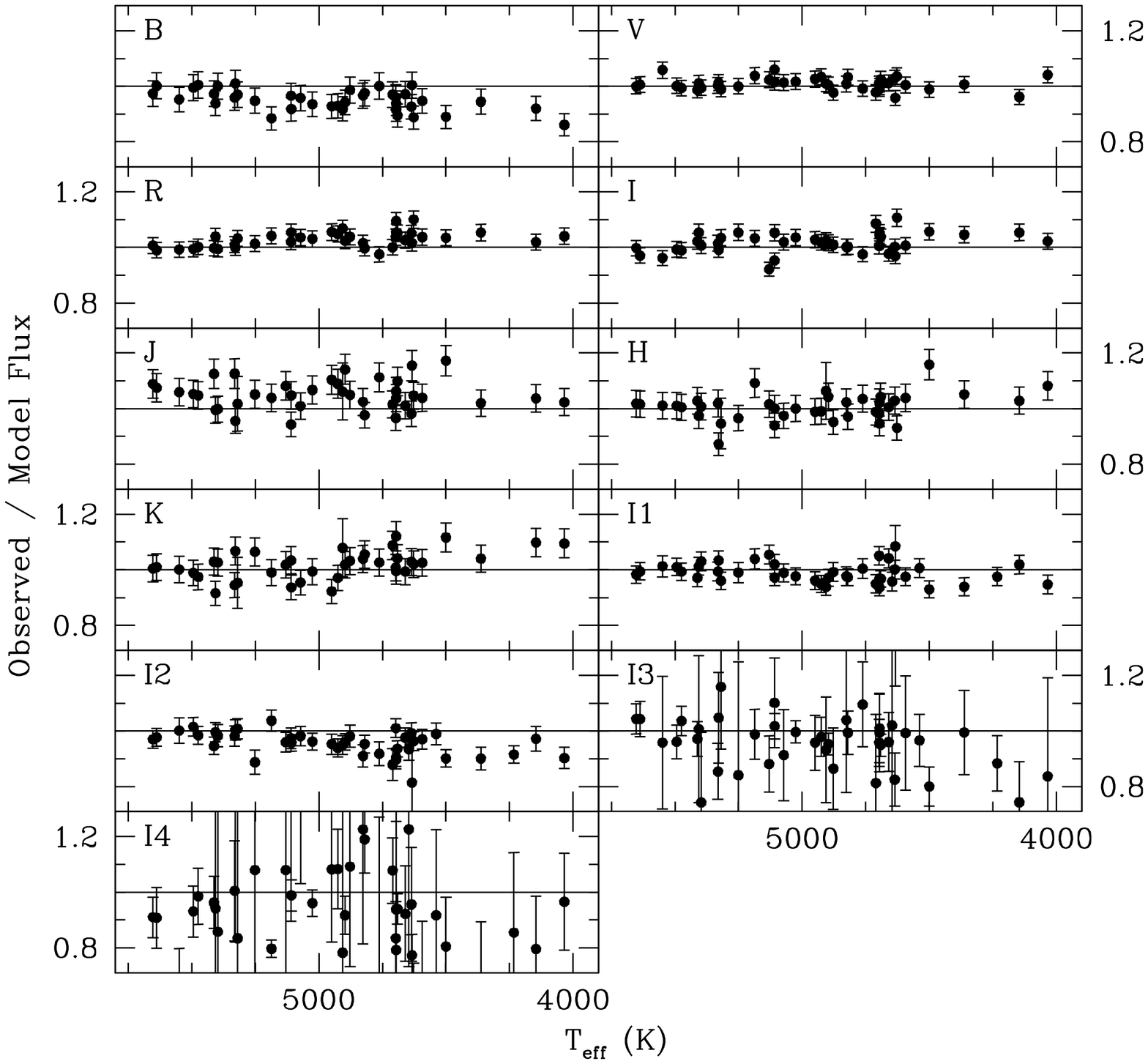}
\caption{Ratio of the observed vs. model fluxes for the best-fit models.}
\end{figure}

\clearpage
\begin{figure}
\hspace{-0.8in}
\includegraphics[angle=-90,scale=.75]{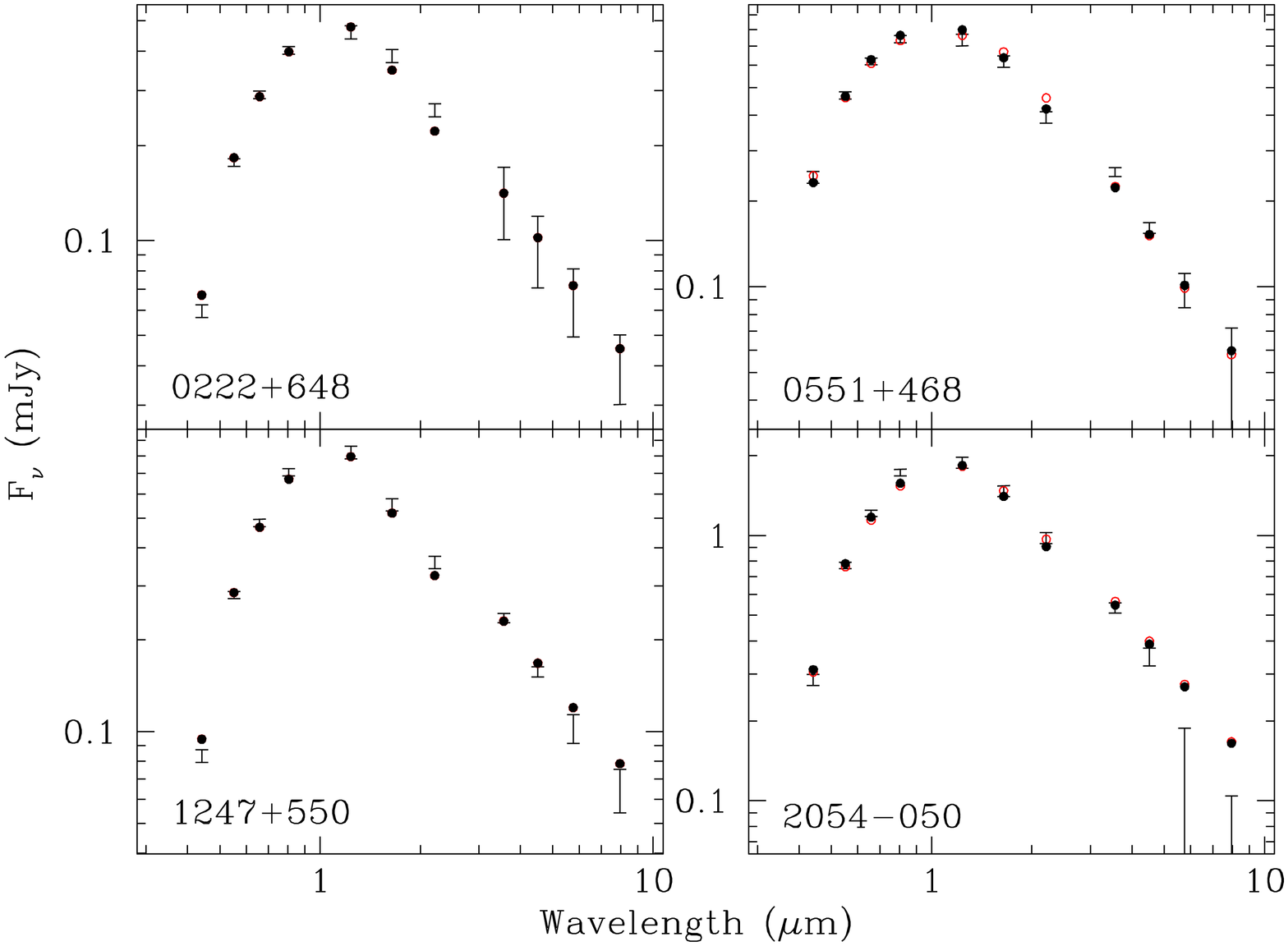}
\caption{SEDs of stars that have nearby bright stars in Spitzer images.}
\end{figure}

\clearpage
\begin{figure}
\hspace{-0.8in}
\includegraphics[angle=-90,scale=.75]{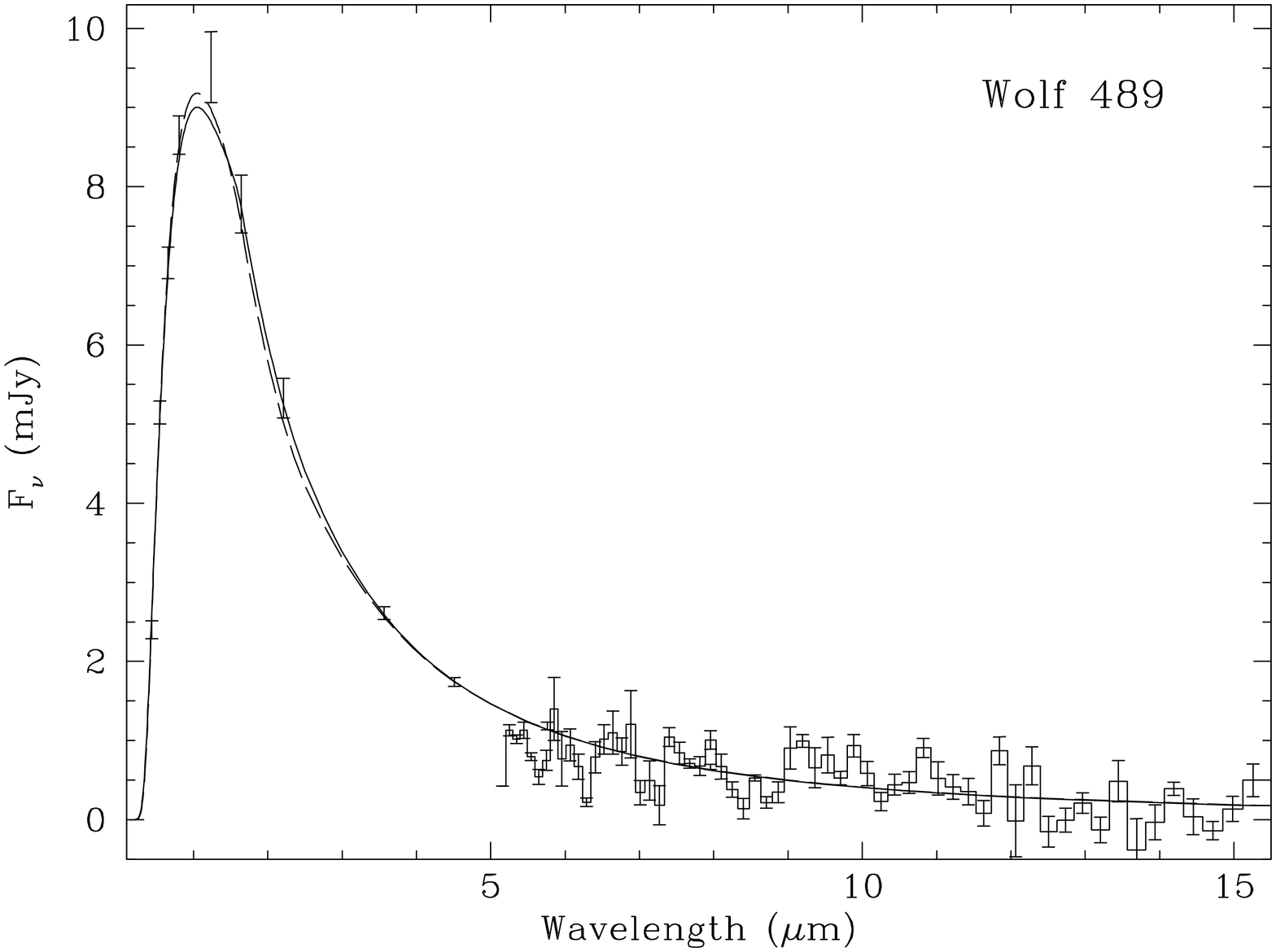}
\caption{The spectral energy distribution of Wolf 489 including the IRS
spectrum from 5 to 15 $\mu$m. The best-fit pure hydrogen and mixed H/He
atmosphere models are shown as solid and dashed lines, respectively.
The IRS spectrum is normalized to match the IRAC photometry in the
8 $\mu$m band.}
\end{figure}

\clearpage
\begin{figure}
\plottwo{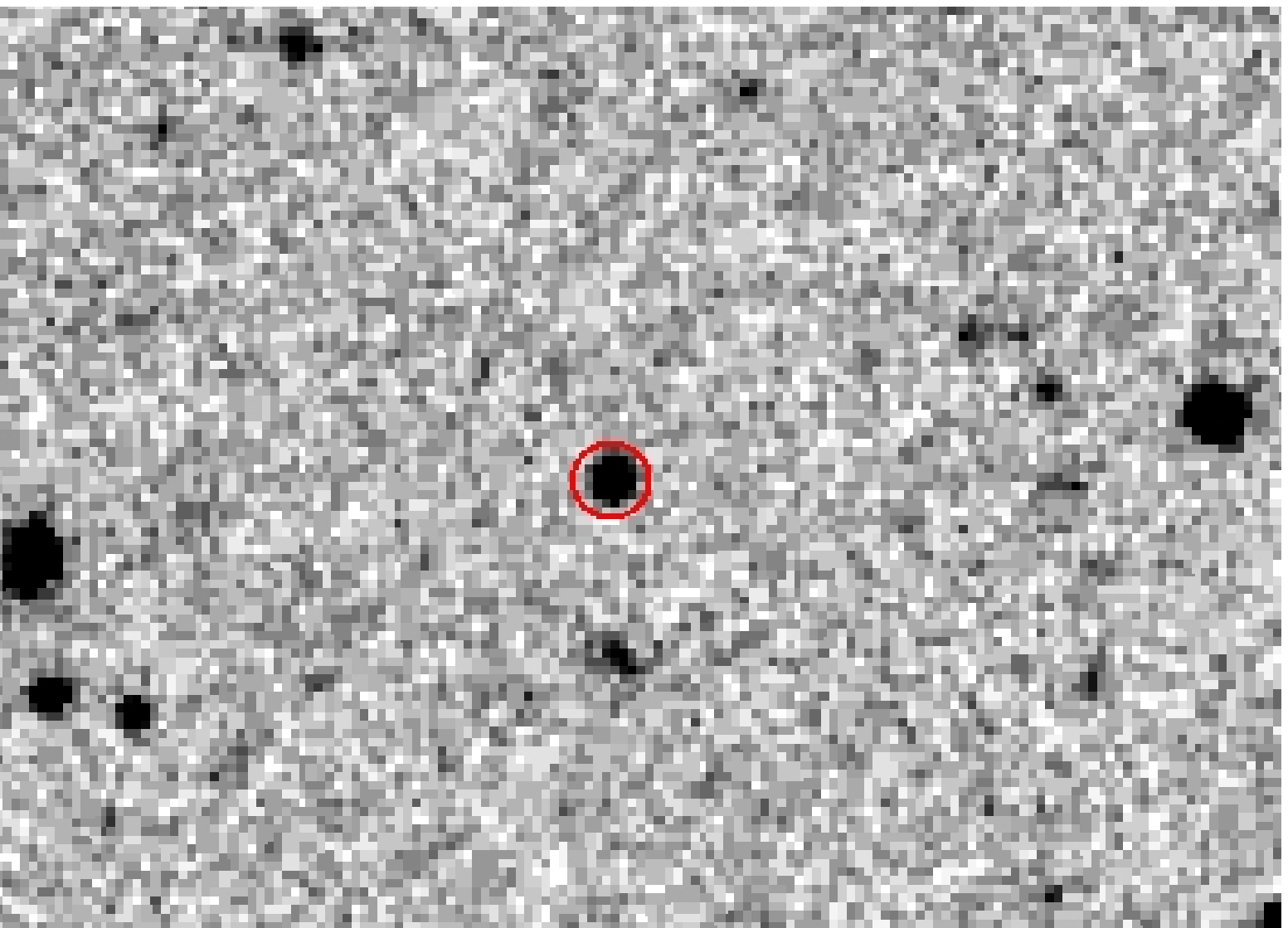}{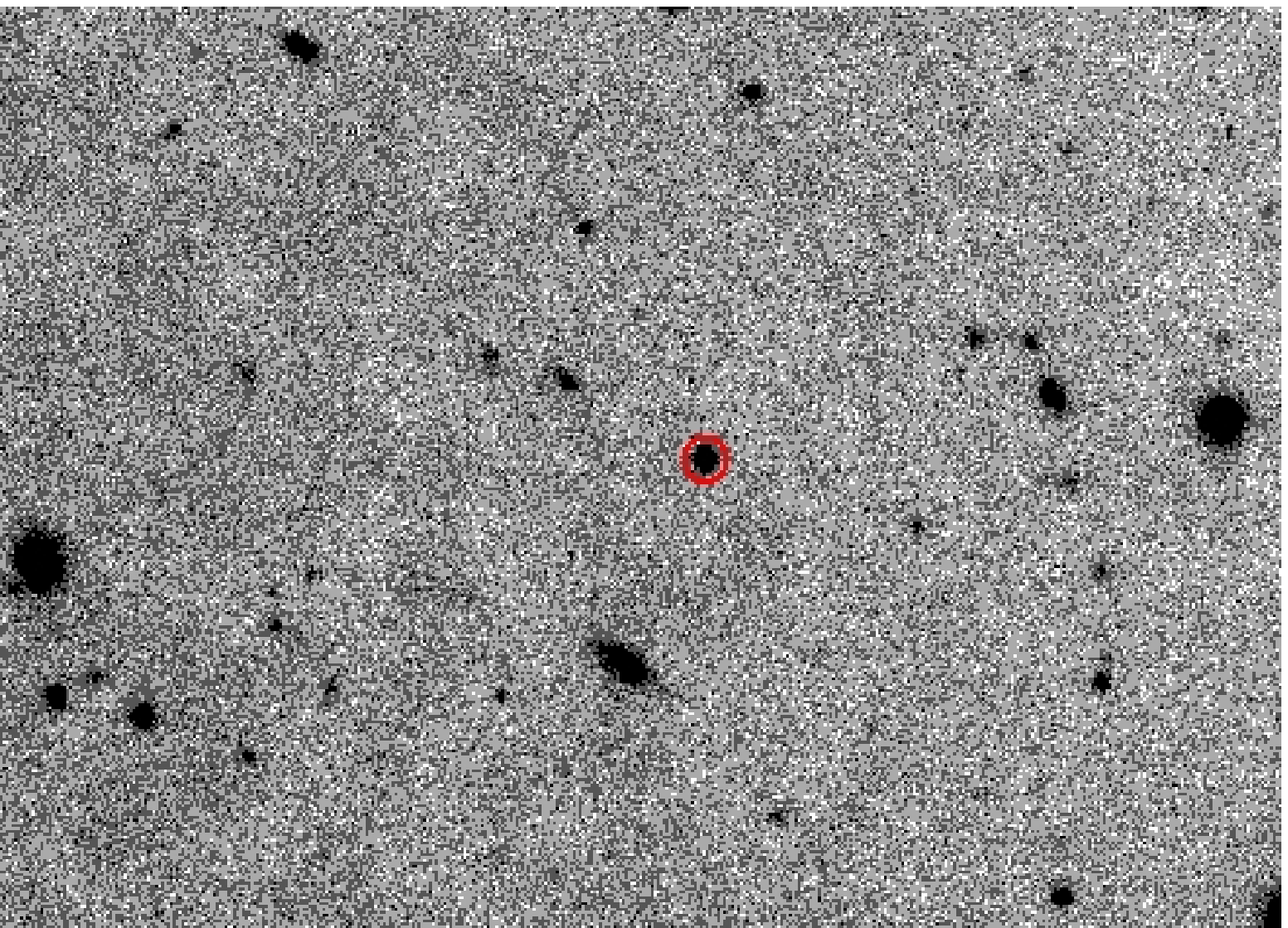}
\plottwo{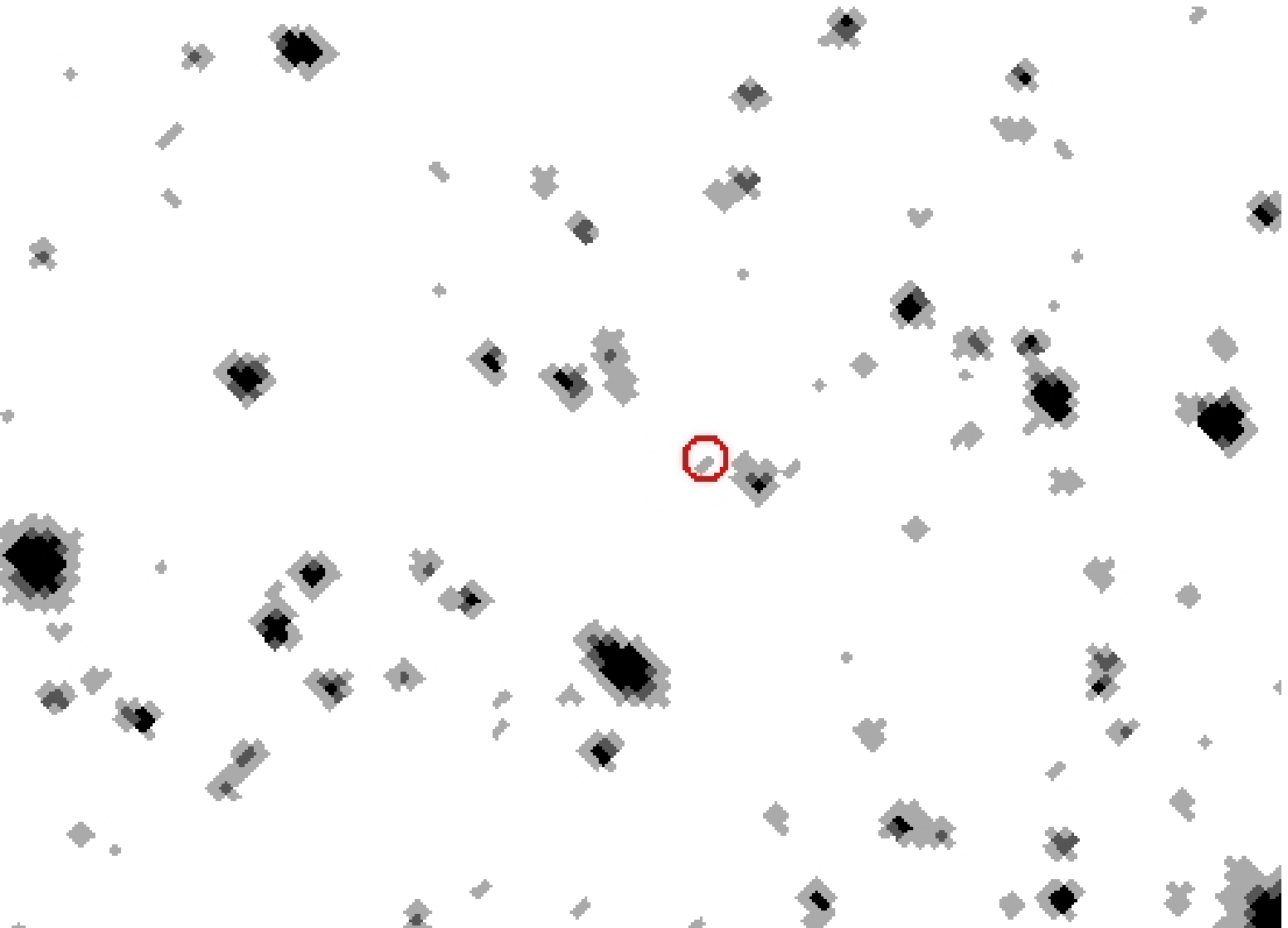}{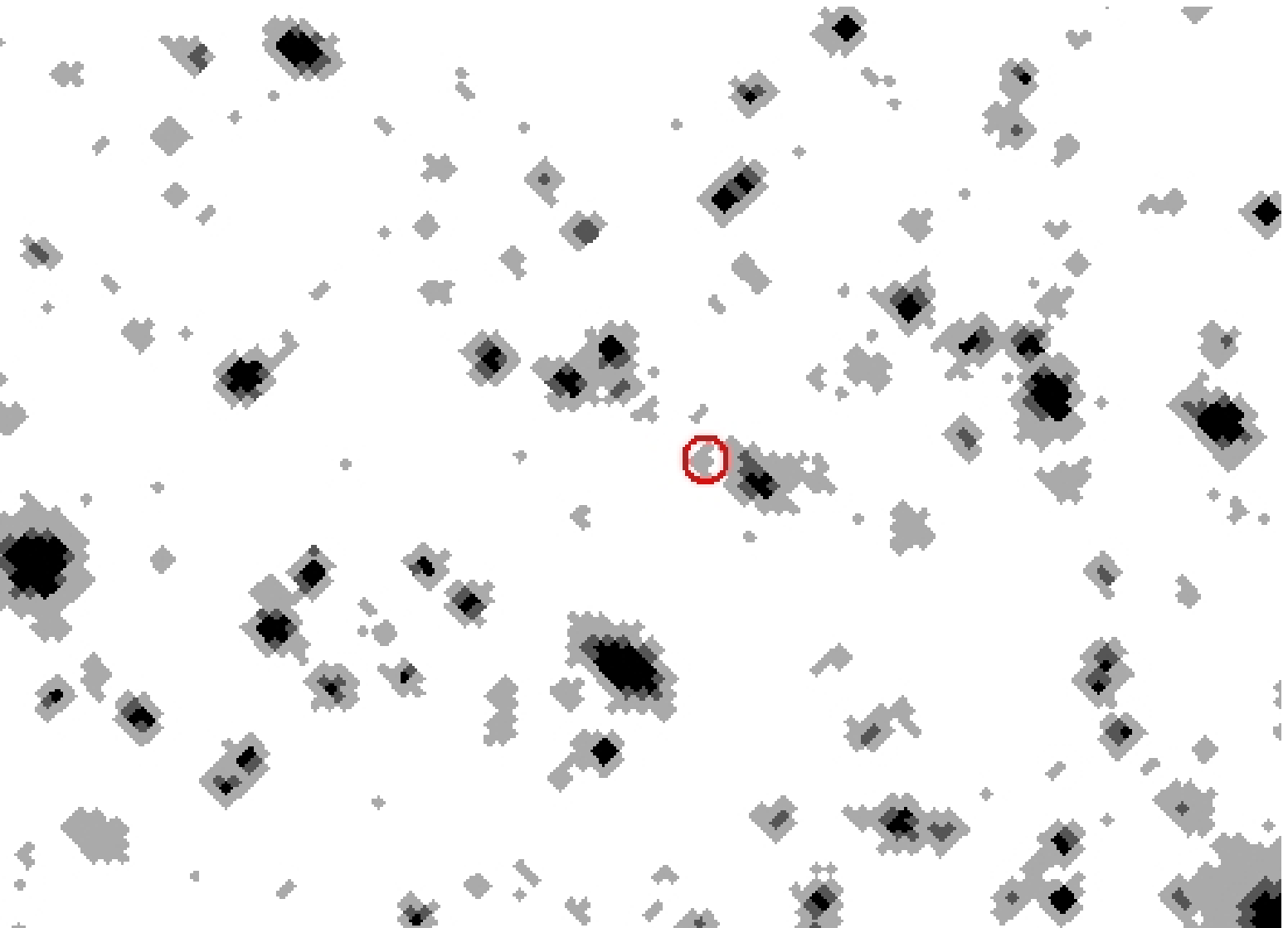}
\caption{LHS 3250 in June 1988 POSSII blue plates with North up and East to the left (top left panel),
May 2007 MDM 2.4m telescope $I-$band image (top right panel),
and August 2006 {\em Spitzer} IRAC Channel 1 and  2 images (bottom left and right panels).}
\end{figure}

\clearpage
\begin{figure}
\plotone{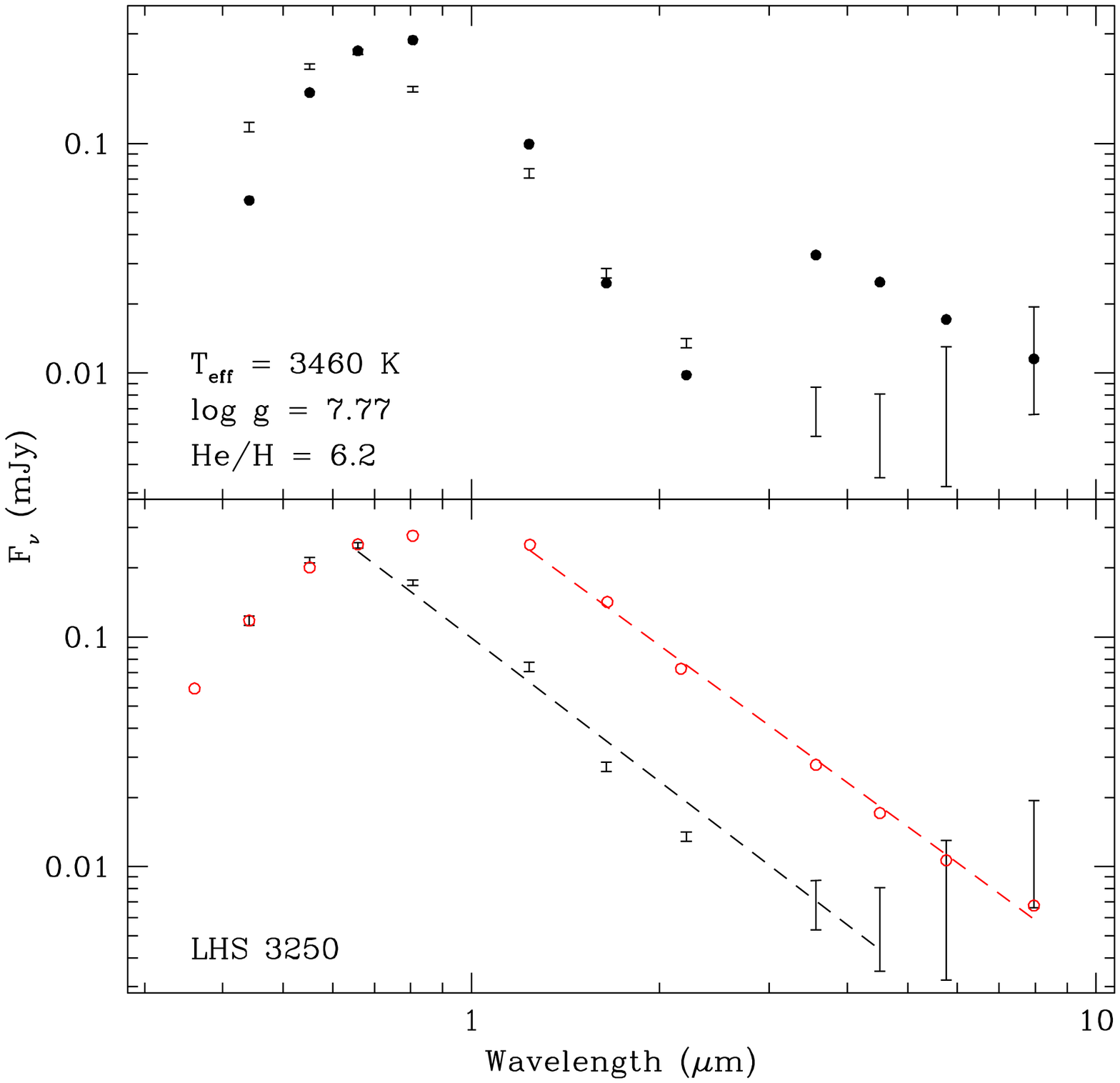}
\caption{The LHS 3250 SED (error bars) compared to the best-fit mixed H/He model atmosphere (filled circles, top panel) and to
LHS 1126 (open circles, bottom panel). The LHS 1126 SED is normalized to match LHS 3250 in the $B-$band.
Power law fits to the infrared portion of the SEDs are shown as dashed lines. The power law fit to the LHS 3250 SED does not include
5.8 and 8 $\mu$m photometry due to large error bars, and it has an index of $-$2.07.}
\end{figure}

\clearpage
\begin{figure}
\hspace{-0.8in}
\includegraphics[angle=-90,scale=.75]{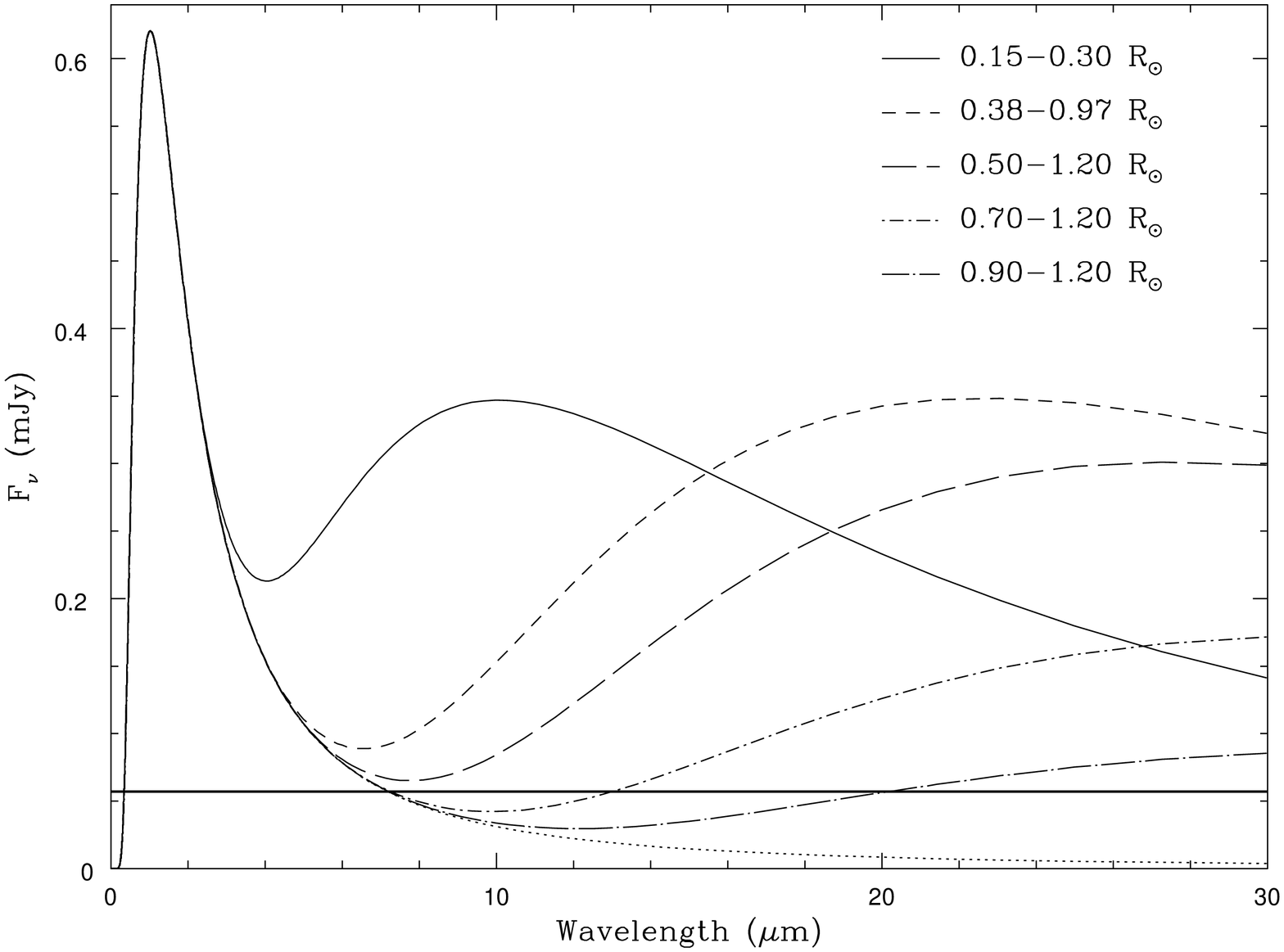}
\caption{The predicted flux distributions of a 5000 K white dwarf (dotted line)
and hypthetical disks at different distances from the star. The inclination angle
is fixed at 45$^o$. The inner and outer radii of the disks are labeled in the figure.
The thick solid line marks 3$\sigma$ detection limit at 8$\mu$m assuming a photometric accuracy of 5\%.}
\end{figure}

\end{document}